\documentclass[useAMS,usenatbib]{mn2e}
\usepackage{epsfig}

\begin{document}

\title[Mass Functions in Clouds and Clusters]{Mass Functions in
Fractal Clouds: The Role of Cloud Structure in the Stellar Initial Mass
Function}

\author[M. Shadmehri and B. G. Elmegreen]{Mohsen
Shadmehri$^{1,2}$\thanks{E-mail:mshadmehri@thphys.nuim.ie; }, Bruce G.
Elmegreen$^{3}$
\thanks{E-mail: bge@us.ibm.com}\\
$^{1}$ Department of Mathematical Physics, National University Ireland, Co Kildare, Maynooth, Ireland\\
$^{2}$ NORDITA, AlbaNova University Center, Roslagstullsbacken 23, SE-10691 Stockholm, Sweden\\
$^{3}$ IBM Research Division, T. J. Watson Research Center, 1101
Kitchawan Road, Yorktown Heights, NY 10598}

\maketitle

\date{Received ______________ / Accepted _________________ }

\begin{abstract}
The possibility that the stellar initial mass function (IMF) arises
mostly from cloud structure is investigated with fractal Brownian
motion (fBm) clouds that have power-law power spectra. An fBm cloud
with a realistic projected power spectrum slope of $\beta=2.8$ is found
to have a mass function for clumps exceeding a threshold density that
is a power-law with a slope of $\alpha=2.35$, the same as in the
Salpeter IMF. Any hierarchically structured cloud has a clump mass
function with about the same slope.  This result implies that turbulent
interstellar clouds produce dense substructure with the observed
pre-stellar core mass function built in from the start. Details of the
clump formation processes are not critical. The conversion of clumps
into stars involves a second step. A one-to-one correspondence between
clump mass and star mass is not necessary to convert the clump mass
spectrum into an IMF with the same power-law slope. As long as clumps
have an internal stellar IMF from sub-fragmentation, protostellar
accretion, coalescence and other processes, and the characteristic mass
for this internal IMF scales with the clump mass, then the IMF slope
above the minimum characteristic mass will equal the clump mass slope.
A detailed review of IMF models illustrates the prominence of cloud
structure as a major component in a wide class of theories. Tests are
proposed to determine the relative importance of cloud structure and
competitive accretion in the IMF.
\end{abstract}

\section{Introduction}

\subsection{Clump Mass Functions and Power Spectra}

Interstellar gas emission has a power-law power spectrum suggesting
structure on a wide range of scales  \citep*{crovisier,green,dickey,miville2003a,miville2003b,miville2010}. Cloud-fitting algorithms that are applied to this
structure find power law mass functions for the emission peaks, which
are usually identified as clouds, clumps, and cores \citep*{williams,stut90}. Because there is considerable
interest in the origin of various interstellar and stellar mass
functions, such as those for giant molecular clouds, molecular cloud
clumps, star clusters, and individual stars, we would like to
understand the relationship between the structure that is viewed as a
featureless power spectrum and the structure that is inferred to be a
collection of discrete objects.

\citet{stut98} suggested that for Gaussian-shaped clouds with
a mass-radius relation $M\propto R^\gamma$ and a power law mass
function of slope $-\alpha$, the slope $-\beta$ of the power spectrum
is given by $\beta=\gamma\left(3-\alpha\right)$. This relation comes
from an integral over the cloud mass function, $dn\propto
M^{-\alpha}dM$, of the contribution to the power spectrum, $dP(k)$,
from a cloud of mass $M$, which is $dP(k)\propto
M^2\exp(-0.5k^2R^2)dn$.  The result is the frequency ($k$) dependent
part of the integral over $M^{2-\alpha} \exp(-0.5k^2R^2)dM$, which is
$k^{-\beta}$ for the above value of $\beta$. \citet{stut98} noted
that the value of $\beta\sim2.8$ they observed corresponded to typical
$\alpha\sim1.6$ and $\gamma\sim2$ for molecular clouds. Observations of
$\beta$ for various regions of the Milky Way are in Table 1; generally
$\beta\sim2.8$.

\begin{table*}\label{beta}
\caption{Observations of Power Spectrum Slope $\beta$ for various
regions of the Milky Way.}
\begin{tabular}{ c c c c c c c c c c c c}
\hline\hline
& Region &&&& Type of Observation &&&&Power Spectrum Slope & Reference\\ &&&&&&&&&($\beta$)\\
\hline
&Foreground of Cas A&&&&HI 21 cm absorption&&&&$2.75 \pm 0.25$&\citet{deshpande}\\
&"&&&&"&&&&$2.86\pm 0.1$& \citet{roy}\\
&Perseus, Taurus, Rosetta clouds&&&&$^{12}$CO&&&&$2.74 \pm 0.08$&\citet{pado2004}\\
&Perseus cloud&&&&$^{13}$CO&&&&$2.86 \pm 0.1$&\citet{pado2006}\\
&"&&&&$^{12}$CO and $^{13}$CO&&&&$\approx 3.1$&\citet{sun2006}\\
&Perseus spiral arm&&&&HI 21 cm&&&&$2.2$ to $3.0$&\citet{green}\\
&Ursa Major high-latitude cirrus&&&&HI 21 cm&&&&$3.6 \pm 0.2$& \citet{miville2003a}\\
&Polaris Flare&&&&$^{12}$CO&&&&$\sim2.8$&\citet{stut98}\\
&"&&&&FIR&&&&$2.7\pm0.1$&\citet{miville2010}\\
&Several molecular clouds&&&&$^{12}$CO and $^{13}$CO&&&&2.5 to
2.8&\citet{bensch}\\
&"&&&&100 $\mu$m&&&&2.9 to 3.2&\citet{gautier}\\
&The Fourth Galactic Quadrant&&&&HI 21 cm&&&&$\sim4$& \citet{dickey}\\
&The Gum nebula&&&&8, 24, and 70 $\mu$m&&&&2.6 to 3.5&\citet{ingalls2004}\\
\hline\hline
\end{tabular}
\end{table*}

\citet{hen2008} derived the cloud mass distribution from
a power spectrum in a different way.  They assumed that the logarithm
of the density, rather than the density, has a power-law power spectrum
(following \citet{beresnyak}), and they considered a log-normal
density pdf, ${\cal P}(\rho,R)$, for scale $R$. Then they equated the
total mass of clouds more massive than $M$ from the integral over the
mass function, $\int_M M^\prime n(M^\prime)dM^\prime$, to the total
mass in the pdf at densities higher than $\rho=M/R^3$, which is $M_{\rm
pdf}(\rho,R)=\int_\rho {\cal P}(\rho^\prime,R) d\rho^\prime$. This
gives $n(M)=M^{-1} dM_{\rm pdf}(\rho,R)/dM= M^{-1} (dR/dM) dM_{\rm
pdf}(\rho,R)/dR$.  The pdf ${\cal P}$ depends on $R$ because the
dispersion $\sigma$ of the log normal function was assumed to depend on
$R$, as
$\sigma^2=\sigma_0^2\left(1-\left(R/L\right)^{n^\prime-3}\right)$ for
$n^\prime>3$. That is, the density fluctuations become smaller when
averaged over larger scales, up to the system scale $L$.  This
dependence comes from the assumption $\sigma^2 \propto
\int_{2\pi/L}^{2\pi/R}k^{-n^\prime}d^3k$ where $n^\prime$ is the
(negative) slope of the power-law power spectrum for $\log(\rho)$. The
result is that $n(M)\propto
M^{-1}\sigma^{-3}(dR/dM)(d\sigma^2/dR)\propto M^{-2}M^{n^\prime/3-1}
=M^{-\alpha}$ for $R<<L$, where $\alpha=3-n^\prime/3$. This is about
the same as in \citet{stut98}, who had $\alpha=3-\beta/3$ for
$M\propto R^3$, considering that $\beta$ in \citet{stut98} was the
power spectrum slope for density while $n^\prime$ in \citet{hen2008} was the power spectrum slope for log-density.

The \citet{hen2008} model has a different relation
between $\alpha$ and $\beta$ when $n^\prime<3$, because then
$\sigma^2=\sigma_0^2\left((L/R)^{3-n^\prime}-1\right)$ diverges as $R$
goes to zero.  When $n^\prime>3$ as in the previous paragraph, $\sigma$
approached a constant for small $R$ and its presence as $\sigma^{-3}$
in the expression for $n(M)$ contributed no additional mass dependence;
only $d\sigma^2/dR$ contributed, making $n(M)\propto
M^{-2}Rd\sigma^2/dR\propto M^{-3+n^\prime/3}$ as above. When
$n^\prime<3$, however, $\sigma$ contains a strong $M$ dependence and
the $\sigma^{-3}$ term is important, as is the $\sigma^2$ in another
term, $(\delta_c+\sigma^2/2)$, which comes from the log-normal density
distribution for threshold density $\delta_c$. This makes $n(M)\propto
M^{-2}\sigma^{-1}Rd\sigma^2/dR\propto M^{-2}\sigma\propto
M^{-2.5+n^\prime/6}$.

\subsection{Clump Mass Functions and Hierarchical Structure}

For hierarchically structured objects, the mass function $n(M)$ always
has a power law slope $\alpha=2$ if the total mass of clouds in a
logarithmic mass interval around a particular mass is the same for all
masses, i.e., for all levels in the hierarchy. Then $MdN(M)/d\log M=$
constant for mass function $N$ in $\log M$ intervals, giving
$N(M)\propto M^{-1}$, or $n(M)\propto M^{-2}$ for linear $M$ intervals.
This $\alpha=2$ result can be derived in many ways. Clouds with an
average of $N$ substructures per structure in all levels $L$ of the
hierarchy have a number of objects per logarithmic mass interval that
increases as $N^L$ and a mass per object that decreases as $1/N^L$,
making $dN/d\log M\propto M^{-1}$ \citep{fleck}. Similarly, the
probability of selecting out of all $\log M$ levels a particular
structure with mass $M$ is proportional to the number $N$ of these
objects, which is also $\propto M^{-1}$ for $\log M$ intervals. The
same result applies to packing scale-free structures in a volume. For
wavenumber $k$ proportional to the inverse of the size, the number of
structures between $k$ and $k+dk$ that fit into a space with dimension
$D$ is $\xi(k)dk\propto k^{D-1}dk$, and the mass of each is $M\propto
k^{-D}$. Converting wavenumber counts into mass counts using the
one-to-one correspondence, $\xi(k)dk=n(M)dM$, we derive $n(M)\propto
k^{D-1} dk/dM\propto k^{2D}\propto M^{-2}$, independent of dimension
$D$. The \citet{hen2008} result for non-gravitating
clouds is shallower than $M^{-2}$ by the index $n^\prime/3-1=2/9$ for
$n^\prime=11/3$ because the total cloud mass in each $\log M$ interval
is not constant with $M$, but decreases with increasing $M$ as the pdf
becomes narrower and the integral of the pdf above a threshold density
becomes smaller.

\subsection{The Stellar Initial Mass Function}

The initial mass function for stars (IMF) seems considerably more
difficult to understand than the mass function for non-gravitating
clouds because the IMF involves the whole star formation process,
including a wide range of densities and diverse physical effects.
Still, the IMF in clusters is often observed to have a power law
component spanning 1.5 to 2 orders of magnitude in the range from
$\sim0.5\;M_\odot$ to $50\;M_\odot$. How much of this power law comes
from a power-law in cloud structure and how much comes from other
processes is not currently known. Current explanations for the IMF
consider cloud fragmentation driven by supersonic turbulence and
self-gravity, clump coalescence, and protostellar accretion \citep[see
reviews in][]{bonnell2007,dib2010,bastian}.

\citet{pado2002} considered an IMF model based on turbulent
fragmentation of a cloud. They proposed that turbulence compresses the
gas into layers of thickness $L\propto V_0^{-1}$ for ambient turbulent
speed $V_0\propto L_0^\eta$, initial length $L_0$, and Kolmogorov
exponent $\eta\sim1/3$. They also assumed that the number of stars
scales with $L_0^{-3}d\log L_0$ from close packing (as above, where we
wrote $\xi(k)dk\propto k^3d\log k$ for $k=1/L_0$ in 3D). Then with
$M\propto \rho L^3=\rho_0L_0L^2$ for cubic regions in the compressed
layer and density $\rho=\rho_0L_0/L$ for uncompressed density $\rho_0$,
it follows that $M\propto L_0^3\left(L/L_0\right)^2\propto
L_0^{3-2\eta}$. Thus $dN(M)/d\log M\propto L_0^{-3}\propto
M^{3/\left(2\eta-3\right)}\sim M^{-1.3}$ when $\eta=1/3$, which is
close to the Salpeter IMF, $dN/d\log M\propto M^{-1.35}$. The final IMF
was assumed to be this clump mass function multiplied by the
probability that a clump of mass $M$ exceeds the thermal Jeans mass;
the pdf for thermal Jeans masses comes from the pdf for density at a
constant temperature. Their clump model steepens the mass function
slope above the value of $-1$ for mass-conserving fragmentation because
they limit the number of stars that can form in each layer of dimension
$L\times L_0\times L_0$ to a constant, even though the available space
scales with $L_0^2$, i.e., $(L_0/L)^2$ cubical volumes fit in this
layer. If each of these cubes were to form a star, then $N(M)$ would be
larger by the factor $(L_0/L)^2$ and we would retrieve the purely
hierarchical result, $dN(M)/d\log M\propto M^{-1}$.

\citet{pado97} and \citet{hen2008} proposed a different model for cloud fragmentation in the
self-gravitating case. They start with the expression written above for
a non self-gravitating cloud, $n(M)\propto M^{-1}dM_{\rm
pdf}(\rho,R)/dM$, but then consider the unstable regions with average
density $\rho$ larger than the density at which the Jeans length equals
$R$. Instead of summing the cloud masses above $\rho R^3$, they sum the
cloud masses smaller than $\rho R^3$, because the cloud's mass will be
less than the Jeans mass at length $R$ when the density exceeds the
Jeans-length density. The result is $n(M)\propto M^{-1}(d\rho/dM)
P_{\rm pdf}(\rho, R)$ where $\rho$ is the density giving $M=M_J$.
\citet{pado97} assume the masses are thermal Jeans masses, in
which case $M\propto \rho^{-0.5}$. \citet{hen2008} consider a turbulent medium, for which $M_J\sim G V^2 R\propto
R^{2\eta+1}$ when $V\propto R^\eta$, and then $\rho\sim M_J/R^3\propto
R^{2\eta-2}\propto M^{(2\eta-2)/(2\eta+1)}$. In both cases, the mass
function is the product of a power law and a log-normal. In part of the
mass range, the slope has the Salpeter value.

\citet{dib2010} presented a semi-analytical model to describe the
simultaneous evolution of the dense core mass function and the IMF,
considering accretion and feedback by winds. The basic core mass
function came from \citet{pado2002}. The results showed mass
functions that varied throughout the cloud and changed with time. The
final functions were those at the time when pre-stellar winds disrupted
the cloud, and they compared well with the core and stellar mass
functions in Orion.  Because the underlying model is the core function
in \citet{pado2002}, geometric effects play an important role in
their IMF.

All of the analytical models just described have an $\alpha=2$ power
law as an underlying mass function, modified in the various cases by
more or less total mass being included in logarithmic intervals as a
function of the cloud mass.  The models that begin with an equation
$Mn(M)=dM_{\rm pdf}(\rho,R)/dM$ for $M_{\rm pdf}$ from the density pdf
effectively assume that an increment toward larger or smaller mass in
the mass function corresponds to an increment in density. For the IMF
models, this means that massive stars correspond to low density clouds
or parts of clouds. This is contrary to the mass segregation that is
often observed in young clusters, and requires accretion or coagulation
as a second step.  The same mass-density relation results from any
isothermal model that identifies star mass with Jeans mass along the
power law part of the IMF. To avoid this contradiction, the pre-stellar
gas temperature has to increase with core mass \citep[e.g.,][]{krumholz}.

\subsection{The Press and Schechter Mass Function}

The cosmological equivalent of these derivations was considered by \citet{press74}. They assumed infinitesimal perturbations in
an expanding universe and determined the mass function of bound objects
as a function of time. The initial density was assumed to have a power
spectrum, $P(k)$, and the distribution function, $\cal P$, for density
perturbation amplitude $\delta$ inside a given scale $R$ was assumed to
be Gaussian in $\delta$ with dispersion $\sigma$ given by
$\sigma^2=(2\pi)^{-3}\int d^3k P(k)W^2(R,k)$. The window function
$W(R,k)$ was introduced by \citet{pad} to avoid a divergence at
high $k$ when $\beta<3$; a Gaussian form
$W\sim\exp\left(-0.5k^2R^2\right)$ was assumed. If $\delta_c$ is the
critical density at the initial time that makes a perturbation bound at
the present time, then the fraction of bound objects at the present
time with masses larger than $M$ is $F(M)=\int_{\delta_c}^\infty {\cal
P} d \delta.$ The co-moving mass distribution function is then
$n(M)=(\rho/M)dF/dM$. This procedure is similar to that followed by \citet{hen2008}, except that Hennebelle \& Chabrier (2008) also
have a lower limit to $k$ in the integral for $\sigma$. This lower
limit introduces a constant part in the expression for $\sigma^2$, and
causes the mass-dependence of $\sigma$ to be negligible when $n^\prime$
($\beta$ in our notation) exceeds 3 in the Hennebelle \& Chabrier (2008)
model. Without a lower limit on $k$, the cosmological mass function is
$n(M)\propto M^{-2}\sigma^{-1}\exp\left(-\delta_c^2/2\sigma^2\right)$,
which is the Press \& Schechter (1974) result. Note the purely geometric
component, $M^{-2}$, as in the other mass function models discussed
above. For initial power spectrum $P(k)\propto k^{-\beta}$ and mass
$M\propto k^{-3}$, the dispersion in the density distribution function
is $\sigma\propto M^{(\beta-3)/6}$. Then $n(M)\propto
M^{-2}M^{(3-\beta)/6}
\exp\left(-0.5\left[M/M_{tr}\right]^{(3-\beta)/3}\right)$. This mass
function is a power law truncated by an exponential. The truncation
mass, $M_{tr}$, is the mass that is at the critical density for
boundedness at the present time. The slope of the power law part,
$\alpha=1.5+\beta/6$, has a positive correlation between $\alpha$ and
$\beta$, unlike the \citet{stut98} and \citet{hen2008} slopes.
We return to this point again in Section \ref{sect:3.2}.

\subsection{The IMF in Numerical Simulations: Competitive Accretion or Cloud Structure?}

Numerical simulations of star formation have a different approach. They
explain the IMF by analogy to the mass distribution of sink particles
formed in the simulation \citep[e.g.,][]{tilley,pado2007,Li2004,nakamura2005,nakamura2007,martel2006,bate2009}. There are many physical processes involved, and it is not generally clear which dominate the IMF. Moreover, the
simulations differ from each other in the presence of magnetic forces,
magnetic diffusion, turbulence driving, initial cloud boundedness, gas
cooling, stellar outflows, starlight heating, and other properties,
even though they get about the same IMF. One thing they have in common
is a turbulent gas, which means hierarchical structure with an $M^{-2}$
underpinning for mass functions. But there are other things in common
too.

It is instructive to see what differs in the few simulations that gave
another IMF. Two produced relatively flat IMFs: one model in \citet*{klessen} that had small-scale turbulence driving, and models in \citet{clark2008} that had transient unbound clouds. What they
had in common was isolated star formation in non-hierarchical clouds,
as determined by a short driving scale in the first case and by rapid
cloud expansion in the second. They also differed from the usual
simulations in having no competition among protostars for the gas: one
sink particle formed in each fragment with an overall efficiency that
was low.

Competitive accretion \citep*{zin82,bonnell2001a,bonnell2001b} involves
cloud mass (and ultimately sink particle mass) that comes from all over
a cloud, independent of the initial mass of the clump that is accreting
\citep[e.g.,][]{clark2006,smith2008}. The final
star mass then depends on the accretion rate integrated over time. This
mass can be small if the star is ejected from its gas reservoir early
\citep{reipurth,bate2002}, or if the accretion is
slowed by any of a number of processes, such as tidal forces in a
cluster \citep{bonnell2008}, low local densities, or
relative motions between the protostar and the gas \citep{clark2009}.
If competitive accretion is dominant over geometric effects, then it
should give the Salpeter IMF in a uniform static cloud that is
artificially seeded with initially identical, moving protostars. The
simulated clouds that get a normal IMF are not uniform like this. They
are always hierarchically structured, or they have a hierarchy of
initially converging velocities that lead to hierarchical fragmentation
on the same time scale as star formation. The sink particles take gas
from these fragments and from the condensing proto-fragments.

The question we ask is whether the mass function of the gas reservoirs
plays an important role in the final mass function of the stars. Even
though accretion is the local process by which stars assemble their
mass, the amount of accreted mass could depend more on the developing
cloud structure (filaments, clumps, etc.) than on the presence of
near-neighbors in a competitive environment.  Everywhere competitive
accretion occurs, cloud fragmentation also occurs at an earlier stage.
Which dominates the power law part of the IMF?

A test for the origin of the IMF in simulations might be to construct
clouds with different density and velocity power spectra and to see if
the resulting slope of the IMF depends on the slope of these spectra,
as in the \citet{stut98} derivation (also found in our models
below). If it does, then cloud structure would seem to have an
important role in the IMF. If it does not, and all density power
spectra lead to the same IMF, then we can probably rule out cloud
structural effects and clump mass functions for the simulated IMFs.

One argument that cloud structure does not play an important role was
given by \citet*{clark2007}. They suggested that if
fragmentation continues during star formation, and if the fragmentation
time is shorter for smaller scales, which is likely, then the IMF from
the time-integral over all cores should be steeper than the
instantaneous core mass function. The observations, however, indicate
that core mass functions have the same slope as the IMF \citep*[e.g.,][]{motte98,rathborne}. Considering time
evolution, the IMF becomes $\propto \omega(k) k^2dk$, where $\omega(k)$
is the formation rate of structures with wavenumber $k$, and $k^2dk$ is
the space-filling partition function discussed above \citep*[e.g.,][]{difazio}. For turbulence, $\omega(k)\sim \sigma(k)k$, where $\sigma(k)$ is
the turbulent speed. Setting $\sigma(k)\propto k^{-\eta}$ for
Kolmogorov turbulence exponent $\eta$, we get $\omega(k)\propto
k^{1-\eta}$. If $M\propto k^{-3}$ for some characteristic density at
collapse, then we get a mass function for linear intervals of mass
$dn(M)/dM\propto k^{3-\eta}dk/dM\propto M^{-2-\left(1-\eta\right)/3}$.
This is the same as the mass function derived above for space filling
structures, but now it has a slope that is steeper by
$\left(1-\eta\right)/3\sim2/9$ for $\eta=1/3$ as a result of the time
dependence. \citet*{palo2007} used this technique to derive the mass
function of fragments in an expanding shell. If the IMF power law is
not steeper by at least $2/9$ compared to the clump mass power law,
then the IMF would seem to have a different origin than the clumps. \citet{dib2010} got around this problem by including accretion onto the
clumps. This shifts the mass function to higher values and offsets the
steepening from the time dependence.

\begin{figure*}
\epsfig{figure=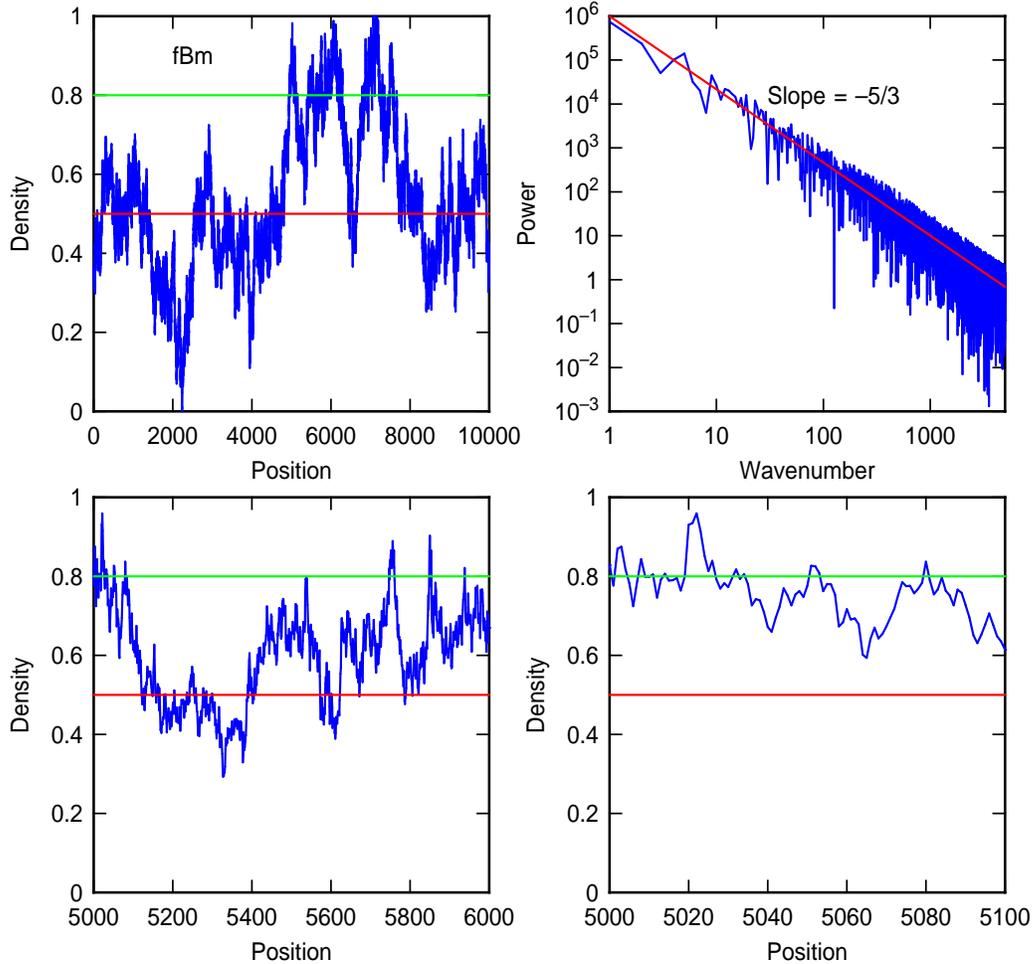,angle=0,scale=0.9} \caption{(Top left)Density scan of a 1D
fBm cloud with two sample thresholds for clump definition. (Top right)
Power spectrum of the density scan. (Lower left) Enlargement of a
section of the fBm cloud, as indicated by the positions on the
abscissa. (Lower right) Greater enlargement of the same cloud.
\label{oneD_example_5over3_1E4long}}
\end{figure*}

\subsection{This Paper: Mass Functions in Fractal Clouds for Clumps
that Exceed a Threshold Density}

Here we are interested more generally in mass functions for
hierarchical and fractal clouds. We determine the mass function for all
clumps denser than a fixed value in a fractal cloud with a given
power-law power spectrum. The dependency of the mass function on the
power spectrum and density threshold are determined. The star formation
process itself is not considered, nor is the IMF below the plateau at
around $0.3\;M_\odot$. Several possible connections between our results
and the full IMF are discussed in Section \ref{sect:disc}.

An approximately fixed density threshold for star formation makes sense
physically \citet*{E2007}. Magnetic diffusion should suddenly become
more rapid at a density of $\sim10^5$ cm$^{-3}$ because the density
scaling for the electron fraction becomes steeper, changing from
$n^{-1/2}$ to $n^{-1}$. This steepening occurs because charge exchange
replaces dissociative recombination for the neutralization of ionic
molecules, and electron recombination on neutral grains replaces
dissociative recombination with ionic molecules \citep[see Fig. 1 in][]{E79,draine}.  If the power of density exceeds $2/3$,
then the time for an initially subcritical core to start collapsing is
comparable to the initial free fall time \citep*{hujeirat}. This is an important condition for star formation.

Dust grains and their coupling to the magnetic field also change at
this density. Small grains normally dominate the viscous cross section
for magnetic diffusion because of their high number density, but PAH
molecules and small grains begin to disappear at densities in excess of
$\sim10^5$ cm$^{-3}$ \citep{boulanger}. Gas depletion at high
density removes ionic metals, and this also lowers the ionization
fraction and promotes enhanced magnetic diffusion.  The depletion time,
$10^{10}/n$ years \citep*{omont}, becomes smaller than the dynamical
time, $\left(G\rho\right)^{-1/2}$, when the density, $n$, exceeds
$3\times10^4$ cm$^{-3}$.  Grain coagulation at this density becomes
important too, and this reduces the number of charged grains and the
net coupling of charged grains to neutrals \citep*{flower}. Further coupling loss arises because
large grains lose their field line attachment at high density \citep*{kamaya}. All of these microscopic effects should significantly
speed up gas collapse at a density near $\sim10^5$ cm$^{-3}$ because
they allow the magnetic field to leave the neutral gas more quickly.
The mass distribution of clumps at this density should then have an
imprint on the final stellar mass distribution.  The final mechanisms
for collapse into stars could be less important to the IMF than the
mass function of clumps at the threshold density where the collapse
begins.

We also study whether higher density limits produce steeper or narrower
mass functions. Such an effect might explain why the power spectrum
slope for the densest cores (which is like the Salpeter IMF slope of
$-2.35$) is steeper than the power spectrum slope for low-density
clouds (which is $\alpha\sim1.6-1.8$). It might also explain the
apparent steepening of the IMF in galactic regions that have generally
low densities, as in low surface brightness spirals \citep{lee2004},
dwarf galaxies \citep{hoversten,hunter}, and
the outer regions of spirals \citep{meurer96,lee2009}. The
IMF would get steeper in these regions if stellar mass is related to
clump mass, if there is a characteristic density at which stars form,
and if the clump mass function gets steeper at higher relative
densities.

We previously found such a correlation between the slope of the mass
function and the threshold density for clump recognition \citep{E2002,EE2006}. Both  studies used three-dimensional
fractal Brownian motion (fBm) clouds with log-normal density pdfs. Here
we experiment with fBm clouds again, first in one-dimension where the
counting statistics are extremely good, and then in three dimensions,
which is more relevant to star formation. Interstellar clouds are not
the same as fBm clouds, but the geometric resemblance is not bad if we
want to address only simple questions. For example, fractal Brownian
motion clouds can be made with the same power spectrum as interstellar
clouds, and they can have the same density pdf. In MHD turbulence
simulations, \citet{dib2008} also found a steepening of the mass
function at higher gas density, although they commented that this
steepening was the result of stronger self-gravity at higher densities.
Our experiments have no self-gravity and are relevant only to the
geometric aspects of cloud structure.

The dependence of the mass function slope $\alpha$ on the power
spectrum slope $\beta$ is considered in detail.  Generally we expect
steeper power spectra to correspond to shallower mass functions. This
is because steeper power spectra have relatively more structure on
larger scales, which means larger masses, so the corresponding mass
function has relatively more high mass clumps.  Clumps are defined
differently here than in \citet*{press74} or \citet*{hen2008}. Here a clump has every part of it above a threshold
density and the entire matter inside the clump boundary is counted in
the clump mass. In Press \& Schechter and in Hennebelle \& Chabrier,
the average density of a clump is above the threshold but smaller
regions inside can be below the threshold. The implications of this
difference are discussed in section \ref{sect:3.2}.

\begin{figure*}
\epsfig{figure=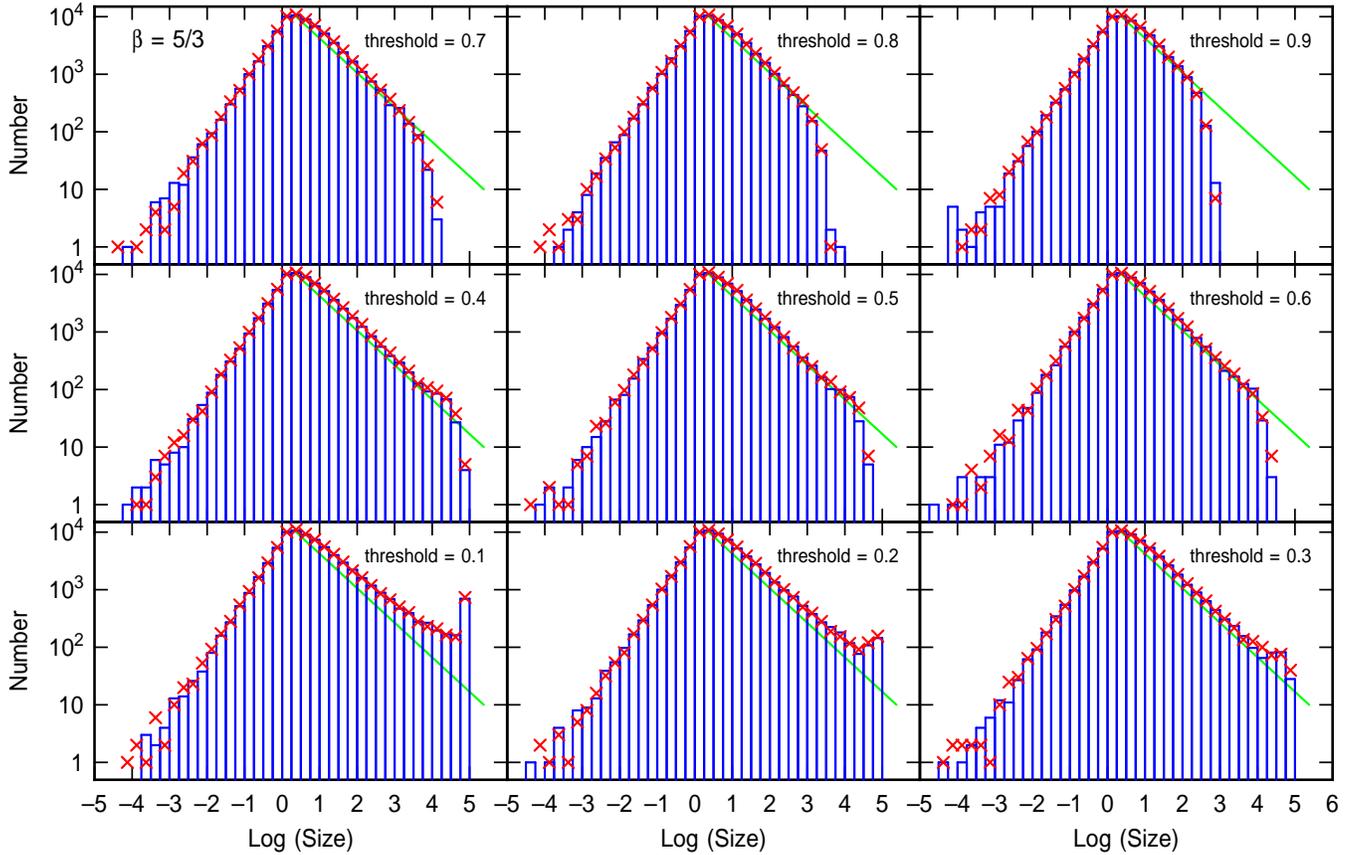,angle=0,scale=0.9}\caption{Size spectra of a 1D
fBm cloud for different density thresholds. The blue histograms are for
the sizes of clumps exceeding the density threshold $C$ and the red
crosses are for the sizes of interclump regions with densities less
than $1-C$. The slope is independent of density threshold at low size,
but the maximum clump size depends on this threshold. The power
spectrum slope is $\beta=5/3$ for all thresholds.
\label{oneDhis2_5over3_1E4high_1E5long_final_3x3}}
\end{figure*}

\citet*{henriksen} also suggested that the IMF comes from counting clumps
in a fractal cloud. In his model, the number of clumps per $\log M$
interval is $N(M)\propto L^{-D}$ for fractal dimension $D$ and scale
$L$. Combined with a density-size relation, $\rho\propto L^{-D_\rho}$,
and a definition of mass, $M=\rho L^D$, the result is $N(M)\propto
M^{-D/\left(D-D_\rho\right)}$. For constant density ($D_\rho=0$), the
geometric result $N(M)\propto M^{-1}$ is recovered. \citet{larson82,larson91,larson92} also modeled the IMF for a fractal cloud but assumed $M\propto L$
for linear gas accumulations along filaments; then $N(M)\propto
M^{-D}$. Our model is similar to Henriksen's, but we characterize
clouds by their power spectrum slope instead of the fractal dimension,
and we consider a fixed density threshold, which makes $M$ effectively
proportional to $L^3$.

Larson (1973) introduced a slightly different fractal model in which
random hierarchical cloud fragmentation produces a log-normal IMF. This
model does not have a power-law component because the stellar masses
are all at the bottom of the hierarchy, in the same level, and their
distribution comes from a spread in the relative number of clumps
versus interclump regions that contribute to the final stellar mass
during the fragmentation process. Power laws tend to arise when only
the fragments can fragment, and log-normals arise when both the
fragments and the inter-fragment gas can fragment during a hierarchical
process of star formation \citep*{E85}. Detailed models of the
first type were in \citet{E97}.  The present model differs from
both of these because we consider only the mass distribution of clumps
denser than a fixed threshold, regardless of fragmentation history or
level in the hierarchy. We assume no particular fragmentation model at
all, only clouds with power-law power spectra.

\subsection{Pre-stellar Clumps}

Pre-stellar clumps have a density close to the proposed threshold
for star formation, and they have about the same mass function as
stars. We suggest that pre-stellar clumps get their mass function from
fractal cloud structure as shown by our models here. The correspondence
between pre-stellar clumps and individual stars is complicated and
discussed in Section \ref{sect:disc}. There is not likely to be a
one-to-one correspondence between dense clump mass and stellar mass,
because pre-stellar clumps can fragment. Still, the full IMF, including
the plateau and the turnover at low mass, might be explained as a
result of a combination of processes with cloud structure the main
contributor in the power-law part.

The mass function for lower-density clumps, like CO clumps in
non-star-forming parts of molecular clouds, are not expected to have
the same slope as the dense clumps studied here. CO requires a high
column density to form because of molecular line shielding and dust
opacity, and it requires a moderately high density for excitation,
although CO line emission is possible even at lower densities if the CO
line opacity is high. Thus CO clumps should not depend so uniquely on a
threshold density as star-forming clumps. With a column density
dependence in addition to a density dependence, CO clumps should have a
shallower mass function than clumps that are defined entirely by
density.  This is because massive low-density clumps that would not be
included in the mass function with a pure density threshold can be
included when there is a column density threshold in addition to a
lower and more fuzzy density threshold.

In what follows, Section \ref{sect:1d} shows the size distribution
functions for clumps in one-dimensional fBm clouds with various power
spectrum slopes and threshold densities. Section \ref{sect:3d} does the
same for three-dimensional fBm clouds in cases with Gaussian density
pdfs and lognormal density pdfs. A discussion of the possible relevance
of these results and other models linking cloud structure to the IMF is
in section \ref{sect:disc}. It is difficult to make a one-to-one
connection between clump mass functions and stellar mass functions
because of certain logical inconsistencies. An IMF that arises from a
combination of cloud structure and late-state accretion seems to be
preferred.  In our model, most of the slope in the Salpeter IMF is from
the universal geometric structure of turbulent clouds.

\section{One-Dimensional Cloud Models}\label{sect:1d}

A one dimensional (1D) density strip with a pre-determined power
spectrum was made by filling a 1D strip in Fourier space with $(0,1)$
noise multiplied by $k^{-\beta/2}$. After Fourier transforming this
$k$-space strip, the result in real space is a density strip with a
power spectrum having a slope $-\beta$. In most of the discussion
below, we normalized each density strip to have a minimum value of 0
and a maximum value of 1; in some discussions, we normalize them to
have a minimum of 0 and an average of 0.5. To make an analogy with
cloud mass spectra, we determined the size spectra of the high density
regions, i.e., above a fixed threshold density. We also determined the
size spectra of the low density regions and of both high and low
density regions together. The size distribution is a good
representation of the mass distribution because most of a cloud's mass
is at a density close to the threshold value \citep[][see also
Sect. \ref{sect:3d} below]{E2002}.

\begin{figure}
\epsfig{figure=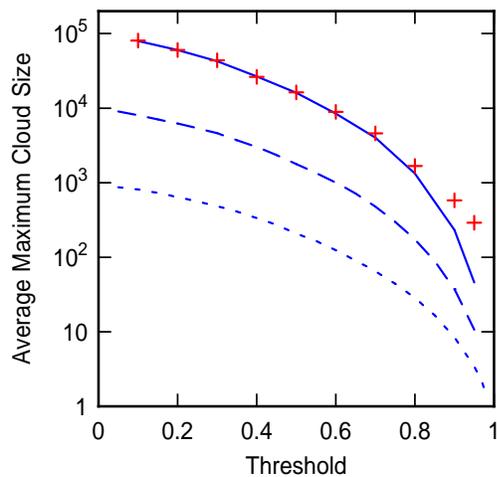,angle=0,scale=0.9}\caption{The maximum clump size versus
the threshold density used to define a clump. The three lines are for
three different total lengths of a 1D fBm cloud.
\label{oneDmaxlength_1E4high_various_lengths}}
\end{figure}
\begin{figure*}
\epsfig{figure=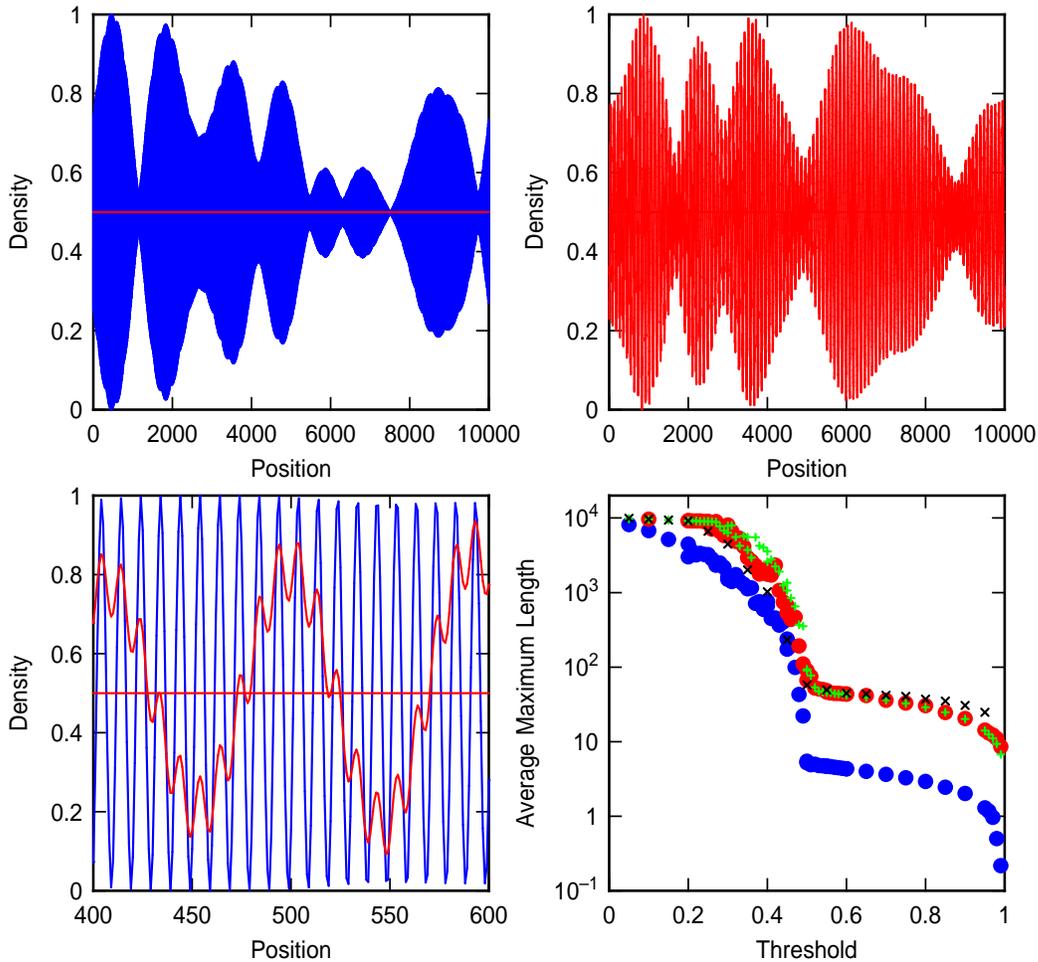,angle=0,scale=0.9} \caption{(Top) Two sample
density scans with restricted wavenumber ranges. (Lower left) The two
scans enlarged and superposed with the same colors as in the top
panels. (Lower right) The maximum clump size versus the threshold
density for the blue and red scans, in the same colors, and for other
composite wavenumbers in black and green. \label{oneD4}}
\end{figure*}

Figure \ref{oneD_example_5over3_1E4long} shows an example. In the top
left panel is a random density strip $10^4$ pixels long that was made
with $\beta=5/3$. The power spectrum of this density strip is in the
top right panel, showing the agreement between the slope and the
expected $-5/3$ slope. The bottom two panels show blow-up sections of
the density strip. A red line indicates a threshold density level of
0.5, and a green line shows the level 0.8. We define a cloud as a
region where the density is above a particular threshold level, so
there will be one set of clouds with a threshold level of 0.5, and
another, smaller set with a threshold of 0.8, etc..  The size of a
cloud or intercloud region is the distance between the two points where
the density profile meets the threshold value. Clearly there are both
large and small clouds at all threshold levels, but there are far fewer
clouds that stick above a high threshold level than above a low
threshold level.

\begin{figure*}
\epsfig{figure=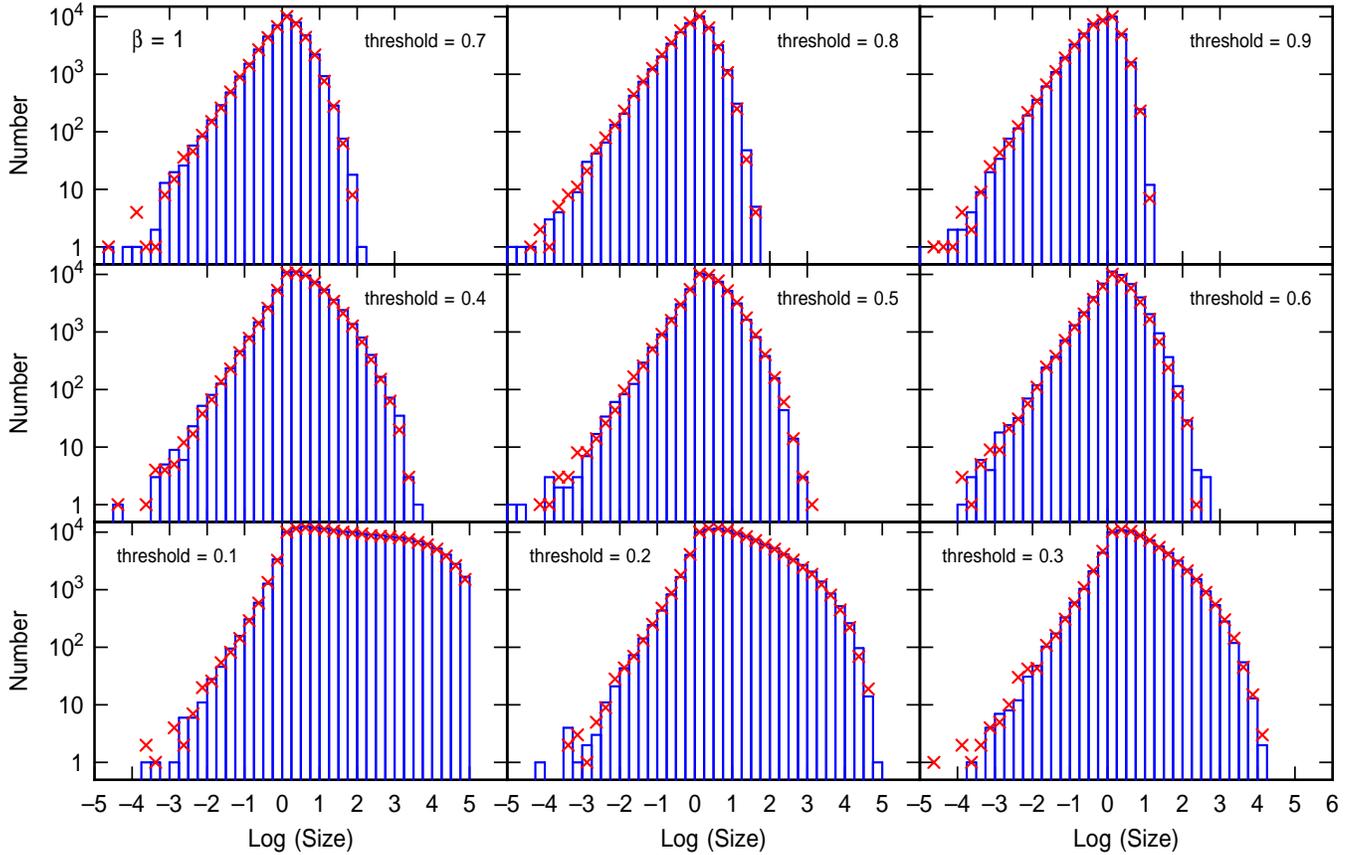,angle=0,scale=0.9}\caption{Mass spectra of clumps
in 1D fBm clouds with power spectrum slope $\beta=1$ for 8 threshold
levels. As in Fig. 2, the blue histograms are for clumps with
thresholds $C$ and the red crosses are for interclump regions with
thresholds $1-C$. \label{oneDhis2_3over3_1E4high_1E5long_3x3}}
\end{figure*}

\begin{figure*}
\epsfig{figure=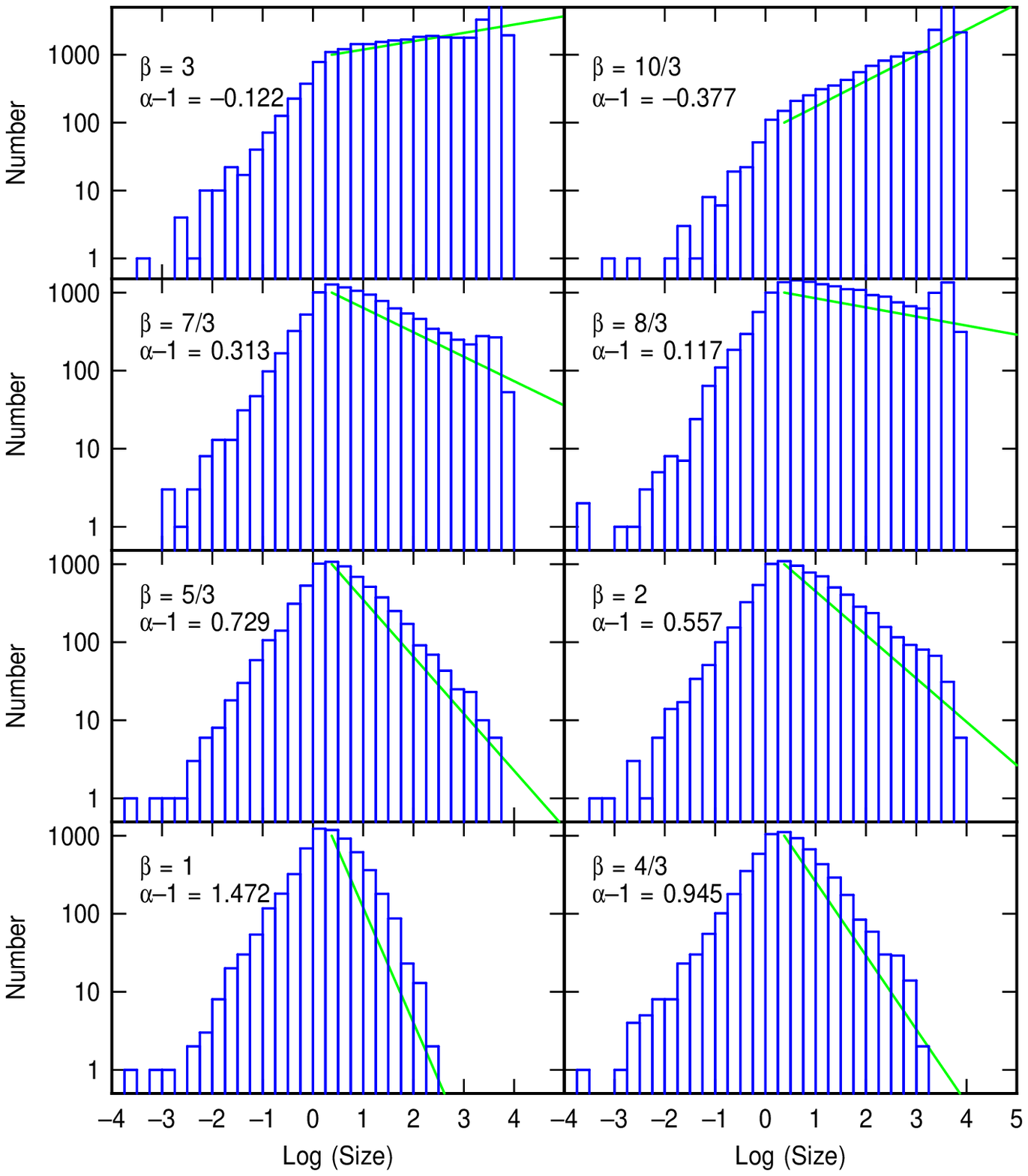,angle=0,scale=0.9}\caption{Mass spectra of clumps in 1D
fBm clouds with various power spectrum slopes, $\beta$, and a threshold
density of 0.5. The slopes of the histograms are indicated by the
values of $\alpha-1$ and by the green lines.
\label{oneDhis2_all_nodes_versusbeta}}
\end{figure*}

Figure \ref{oneDhis2_5over3_1E4high_1E5long_final_3x3} shows histograms
in blue of the size distribution of clouds for various threshold
levels, $C$, and it also shows histogram values, plotted as red
crosses, for the size distribution of low-density regions for the
corresponding threshold values $1-C$. That is, the cloud size
distribution for $C$ is shown in the same panel as the inter-cloud size
distribution for threshold $1-C$. Values of $C$ are given in the
panels. The clouds and the inter-cloud regions mirror each other in
size distributions. This is a sensible result as the density strip is
statistically symmetric with respect to inversion from top to bottom.
The histograms were made for 1D fBm density strips having $10^5$
pixels, and enough strips were made for each $C$ that the number of
clouds or inter-cloud regions at a size in pixels with a logarithm
between 0 and 0.25 was $10^4$ to maintain good counting statistics. In
each case, $\beta=5/3$. The green line in the upper mass range has a
slope of $-0.6$, for comparison purposes. The rising part of each curve
is for sub-pixel sizes.

The maximum size of a cloud decreases with increasing threshold
density. Higher thresholds have smaller clouds and larger intercloud
regions. Figure \ref{oneDmaxlength_1E4high_various_lengths} shows the
trend of average maximum cloud size per simulation versus the threshold
level. The solid curve is for fBm strips of $10^5$ pixels length, the
dashed curve is for strips with $10^4$ pixels, and the dotted curve is
for strips with $10^3$ pixels.  These are for densities that are
normalized to have a minimum value of 0 and a maximum of 1. For each
curve, the maximum cloud size is essentially the whole strip length
when the threshold is low, and the maximum cloud size decreases to zero
when the threshold is large. The red plus symbols are for a second case
where the density strips are normalized to a minimum of 0 and an
average of 0.5, with an arbitrary maximum.  The maximum cloud sizes are
about the same as in the $(0,1)$ normalization except at high
thresholds, where the limitless maximum still produces high density
regions.

The variation in average maximum length with threshold level may be
understood from Fourier analysis. Each density strip is the sum of
sines and cosines for a wide range of wavenumbers and for amplitudes
that vary with wavenumber as $k^{-\beta/2}$ with $\beta=5/3$ in the
cases shown. The short wavelength oscillations modulate the long
wavelength oscillations, and so the high intensity part of any long
wavelength oscillation has ripples from the short wavelength
oscillations. Figure \ref{oneD4} shows an example. The top two panels
show the density for the full lengths of $10^4$-pixel strips with
different Fourier components. The bottom left panel shows the same two
strips with the same color coding magnified in the pixel range from 400
to 600. The bottom right panel plots the maximum cloud size versus the
threshold level, as in Figure
\ref{oneDmaxlength_1E4high_various_lengths}, for these two strips. The
blue strip in the top and bottom left has wavenumbers only between 1000
and 1010, i.e., it occupies a narrow range in Fourier space with the
other Fourier components set to zero. In real space this strip consists
of nearly pure oscillations that each go through the midpoint value of
0.5. The range of wavenumbers modulates the amplitude of this
oscillation (top left panel) with a beat frequency from the difference
in wavenumbers. In this case, the maximum width of a cloud (lower right
panel, blue dots) is nearly independent of the threshold level for
thresholds larger than 0.5 because each cloud is nearly a pure sine
wave. Below a threshold of 0.5, the length of a cloud is the length of
the lower envelope of beat frequency oscillation, which is nearly the
full strip length, $10^4$.

The red strip in the top right and bottom left of Figure \ref{oneD4}
has wavenumbers between 100 and 110 and between 1000 and 1010. There is
still the high frequency oscillation and modulation of this oscillation
as in the blue strip, because the same high frequencies are present,
but there is also a low frequency modulation that pushes up and down
the high frequency modulation (lower left panel). This low frequency
modulation also has a varying amplitude with a beat frequency equal to
the difference in wavenumbers (also 10 wavenumbers).  Thresholds below
0.5 (lower right panel, red dots) have the same long maximum length as
in the blue strip, from the beat frequency, and above this threshold
the short length oscillations are seen again.  The maximum cloud length
at high threshold is higher for the red strip than for the blue strip
because the minimum frequency is lower for the red strip. The high
threshold lengths are about equal to the number of pixels in the strip,
$10^4$, divided by $2\pi$ times the minimum $k$ value. For the blue
strip, this is $10/2\pi$, or around 2, and for the red strip, this is
$100/2\pi$, or around 20.

The green plus symbols in the lower right panel of Figure \ref{oneD4}
are for a case where the density strips have wavenumbers only in the
range of $100-110$. The black crosses are for a case where the strips
have wavenumbers of $100-110$, $310-320$, and $1000-1010$. These two
case have about the same maximum cloud lengths versus threshold levels
as the red case because the maximum cloud length at high threshold is
dominated by the lowest frequencies, which are the same in each case.

The lower right panel of Figure \ref{oneD4} resembles the lower panel
of Figure \ref{oneDmaxlength_1E4high_various_lengths}, illustrating
that the maximum length decreases with increasing threshold because of
the isolation of higher frequencies at higher thresholds. The low
frequency structure is dominated by beating effects between neighboring
frequency groups, and these effects determine the large-scale envelope
of the density strip. This is how a cloud length can be greater than
half the maximum wavelength in a strip.


Cloud size distributions were also determined for other power spectrum
indices using the density normalization from 0 to 1. Figure
\ref{oneDhis2_3over3_1E4high_1E5long_3x3} shows the case $\beta=1$ for
a range of threshold densities, and Figure
\ref{oneDhis2_all_nodes_versusbeta} shows histograms for many $\beta$,
all with a threshold density of 0.5. The slopes of the high mass parts
of the mass functions are indicated in Figure
\ref{oneDhis2_all_nodes_versusbeta} in the upper left corner of each
panel and by the green line. These are slopes on the $\log-\log$ plots,
and therefore equal to $\alpha-1$. The mass function slope gets more
negative as the power spectrum slope gets less negative. This is
sensible because more negative mass functions correspond to relatively
more low mass clouds, and this corresponds to relatively more high
frequency power in Fourier space.

\begin{figure}
\epsfig{figure=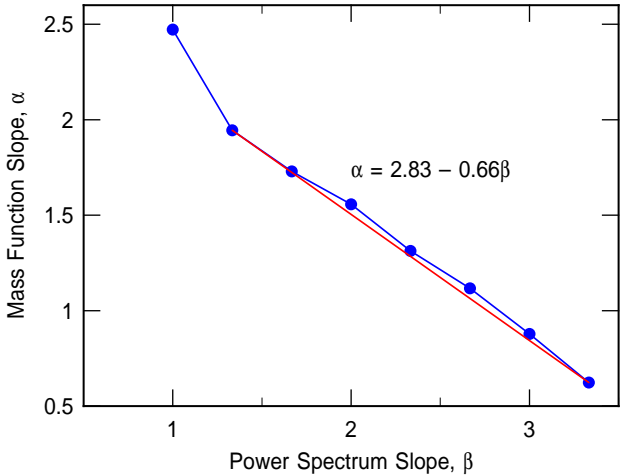,angle=0,scale=0.7}\caption{The slope of the clump mass
function versus the slope of the power spectrum for 1D fBm clouds with
a threshold for clump definition equal to 0.5.
\label{oneD_slope_vs_beta}}
\end{figure}

Figure \ref{oneD_slope_vs_beta} shows the mass function slopes $\alpha$
versus the power spectrum slopes $\beta$ for the 1D fBm clouds. The
data fit the relation $\alpha=2.83-0.66\beta$ between $\beta=4/3$ and
$11/3$. The threshold value was 0.5 for this. Higher thresholds would
not change the result much because the slope $\alpha$ is independent of
the threshold below the upper clump mass (which does depend on the
threshold). According to \citet{stut98} and \citet{hen2008}, the relation is $\alpha=3-\beta/\gamma$ for $\gamma$
equal to the power in the mass-radius relation. In our 1D case,
$\gamma=1$ and this equation would suggest $\alpha=3-\beta$. Our actual
result is close to this.

\begin{figure}
\epsfig{figure=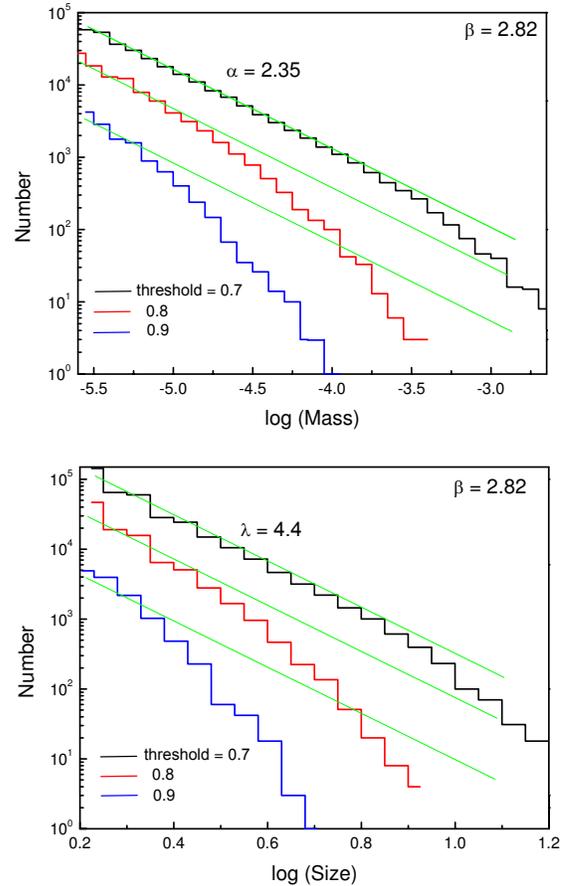,angle=0,scale=0.45}\caption{Mass and size
functions for 3D fBm clouds with Gaussian pdfs. The power spectrum
slope is $2.82$, which produces a mass function similar to the Salpeter
IMF for low and intermediate masses. The drop in the mass spectrum for
high mass and high threshold density is a result of a superposition of
low and high frequency wavenumbers in the fBm cloud, as discussed in
Figure 4. Green lines are fitted to the low mass and size portions of
the distributions.\label{beta2-82final}}
\end{figure}

\section{Three Dimensional fBm Models}\label{sect:3d}

Clump mass functions were also determined for three-dimensional clouds
with and without log-normal density pdfs. The log-normal was made to
resemble isothermal clouds in simulations of MHD turbulence
\citet{vaz}. For 3D, we calculated clump mass in addition
to volume, although the mass turned out to be nearly equal to the
volume multiplied by the threshold density in all cases, as assumed
above for the 1D experiments.

\begin{figure}
\epsfig{figure=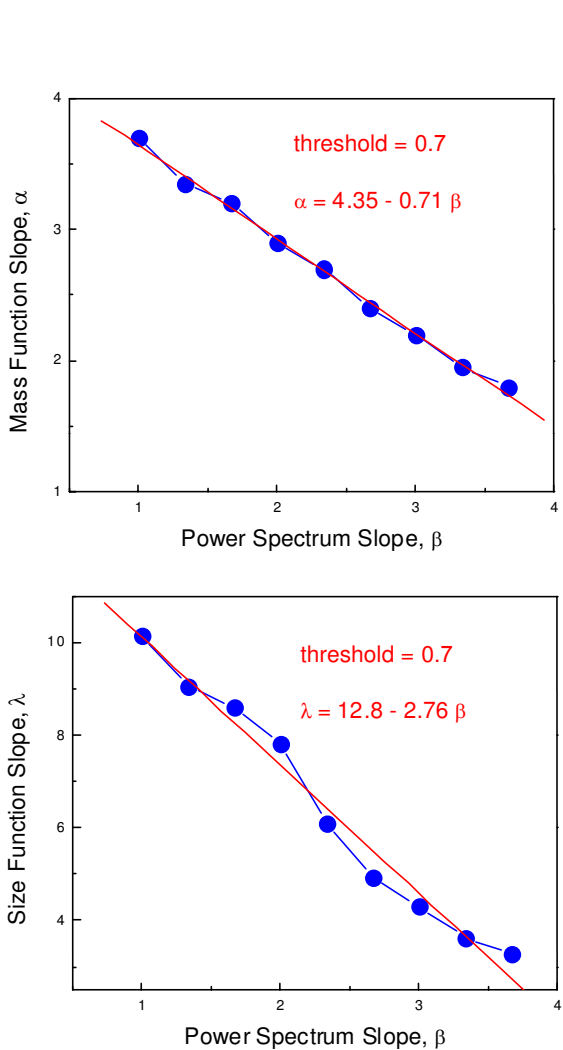,angle=0,scale=0.5} \caption{The slopes of the low
mass and low size parts of the mass and size functions are shown versus
the power spectrum slopes for 3D fBm clouds with a density threshold
for clumps equal to $0.7$. \label{slopes}}
\end{figure}

\begin{figure}
\epsfig{figure=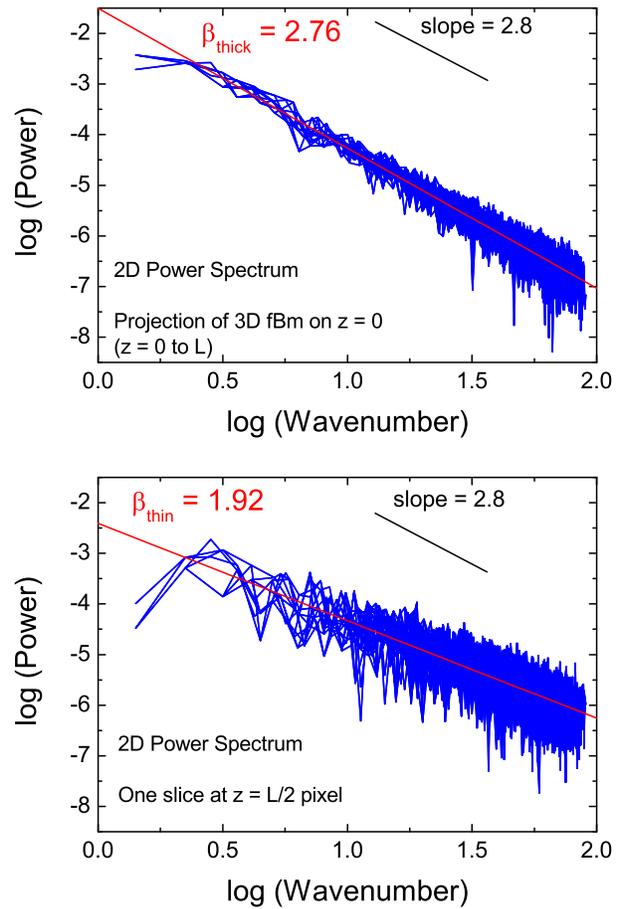,angle=0,scale=1} \caption{(Top) Power spectrum
of a 3D fBm density field made by the integrating the density along the
z axis from $z=0$ to $L$ (the size of the box in pixels). (Bottom)
Power spectrum of the density field of a slice at $z=L/2$ with one
pixel thickness.  The slope of the power spectrum of a 3D fBm density
field is shown at the top of each panel. The projected power spectrum
has about the same slope as the 3D power spectrum, but the slice has a
slope that is shallower by about 1.\label{psslope}}
\end{figure}

The clouds were made in several steps. First we filled a $256^3$ cube
in wavenumber space with random complex numbers, $(k_{x}, k_{y},
k_{z})$, uniformly distributed from $0$ to $1$. Then we multiplied each
number by $k^{-\beta / 2}$, where
$k=(k_{x}^{2}+k_{y}^{2}+k_{z}^{2})^{1/2}$. We took the inverse Fourier
transform of this cube to make a real space density distribution
$\rho_{\rm fBm}(x, y, z)$. This distribution is a standard fBm cloud,
and it has a power spectrum with a negative slope $\beta$. Some of the
clump mass functions presented below use this fBm cloud. To make a
log-normal density pdf, we exponentiated this result:
\begin{equation}\rho_{LN} (x, y, z) = \exp (\rho_{\rm fBm}(x, y,
z)/\rho_{0})\label{eq:rho},\end{equation} where $\rho_{0}$ is for
normalization. To build up the statistics of the cloud mass and size
distributions, we generated an ensemble of $\sim100$ fBm distributions
for each of the three normalizations discussed below. Clumps are
defined as connected pixels above a threshold density inside the
overall cloud.  The clump mass is the sum of the density values in all
of these connected pixels, and the clump size is the cube root of the
number of pixels.

\subsection{fBm clouds with Gaussian density pdfs}

Figure \ref{beta2-82final} shows the mass and size distributions for
clumps in standard fBm clouds with Gaussian density pdfs in the case
where $\beta= 2.82$. This $\beta$ was chosen because it gives a mass
function slope similar to the Salpeter IMF at a cloud-defining minimum
density of 0.7. Three threshold densities are considered. The low mass
and small size parts of the clump functions have about the same slope
for all threshold densities, and there is a steepening of the functions
at intermediate mass and size. This steepening occurs at smaller masses
and sizes when the threshold density is higher, as for the 1D results.

Figure \ref{slopes} shows the slopes of the low-mass parts of the clump
mass and size distribution functions versus the power spectrum slope,
$\beta$, for a density threshold of 0.7. There is a decreasing trend as
for the 1D case. Steeper power spectra produce shallower mass and size
distribution functions. The slope of the Salpeter mass function is
$\alpha=2.35$ on this plot, and that is associated with $\beta=2.82$.
This is the 3D $\beta$ in the cloud volume. The relation between the
two quantities is $\alpha=4.35-0.71\beta$ for a threshold of 0.7.
Recall that the \citet{stut98} and \citet{hen2008} result for 3D with $\gamma=3$ is $\alpha=3-0.33\beta$.

\subsection{Comparison between our Model and the Press-Schechter
or Hennebelle-Chabrier Models}\label{sect:3.2}

The decrease in $\alpha$ with increasing $\beta$ in Figure \ref{slopes}
is opposite to the trend in Press \& Schechter (1974) because of a
difference in the definition of clumps. In Press \& Schechter (1974) and in
Hennebelle \& Chabrier (2008), a clump is defined as an object with an
{\it average} density above a threshold $\rho_c$. The clump mass is the
sum of the masses of only the dense parts ($\rho>\rho_c$) of this
clump. Generally, the clump contains both high density ($\rho>\rho_c$)
and low density ($\rho<\rho_c$) parts, and the high-density mass is
taken equal to the total mass multiplied by the probability that the
density is high. This probability is the integral over the density pdf
for all densities above the threshold. The density pdf is written as
$(2\pi\sigma^2)^{-0.5}\exp(-0.5\delta_c^2/\sigma^2)$ for dispersion
$\sigma^2=(2\pi)^{-3}\int_{k_{min}}^{k_{max}} P(k)dk^3$ and power
spectrum $P(k)\propto k^{-\beta}$ (Padmanabhan 1993). Evidently,
$\sigma\propto k^{(3-\beta)/2}\propto M^{(\beta-3)/6}$, as discussed in
the Introduction.  For $\beta<3$, the integral for $\sigma$ depends
only on the upper boundary, $k_{max}$, and includes all $k<k_{max}$. As
$M$ increases, this upper boundary decreases and the range of $k$
decreases as well.  This is why the dispersion of the pdf, which
depends on the range of $k$, decreases with increasing $M$ when
$\beta<3$. For $\beta>3$, the integral for $\sigma$ depends only on the
lower boundary, $k_{min}$, and includes only $k>k_{min}$ up to some
cutoff in $k$, which may be viewed as the inverse of a smoothing
length.  As $M$ increases, the lower boundary to $k$ decreases,
increasing the range of $k$ between that lower boundary and the upper
cutoff in $k$. As a result, the dispersion $\sigma$ increases --
opposite to the trend for $\beta<3$.

Writing either of these two $k$ limits in terms of $M$, the clump mass
function equals $n(M)\propto
(M^2\sigma)^{-1}\exp(-0.5\delta_c^2/\sigma^2)$, so the power law part
at low $M$ has a slope $\alpha=1.5+\beta/6$. This is the Press \&
Schechter result. The probability that the density is high inside a
region increases with increasing region mass for $\beta<3$ and
decreases with increasing region mass for $\beta>3$, in proportion to
$\sigma^{-1}$ at low $M$ (where the exponential is not important). In
the first case, $\beta<3$, this means that clumpy regions have
dense-gas masses that progressively get smaller than in the purely
hierarchical mass function ($M^{-2}$) as $M$ decreases. As a result,
the clump mass function is shallower than $M^{-2}$ when $\beta<3$ in
the Press \& Schechter formulism. (Visualize an $M^{-2}$ mass function
that shifts to lower $M$ at low $M$ and higher $M$ at high $M$,
stretching horizontally on such a plot.) Similarly in the case
$\beta>3$, clumpy regions have progressively more mass than in the
$M^{-2}$ function as $M$ decreases, making the mass function steeper
(visualize the same initial $M^{-2}$ function, but now compress it
horizontally). This is opposite to the trend in Figure \ref{slopes}.
Physically, what is happening is that the density distribution is
smoother (smaller $\sigma$) at larger $M$ when $\beta<3$, so a region
with average density exceeding some threshold has fewer and smaller
dips to sub-threshold density inside of it. Nearly the entire enclosed
mass is then counted. Smaller mass regions for $\beta<3$ have large
density fluctuations inside of them, so that even when the average
density exceeds the threshold, there are still a lot of regions and a
lot of mass at densities less than the threshold. In this case, the
mass fraction above threshold is small.

Another difference between our model and that of Press \& Schechter
(1974) or Hennebelle \& Chabrier (2008) concerns the lower limit to the
mass. For $\beta<3$, this lower limit is important, as discussed above.
In the Press \& Schechter formulism, the region defined to be a clump
does not depend much on the resolution scale, which is the lower limit
to $M$. Clump regions stay about the same as the resolution increases,
and all that happens is that clump masses get more accurate. That is,
smaller scale structure occurs near the density threshold and the pdf
provides a more detailed representation of the fraction of the mass
above threshold. In our model, the definition of each clump changes as
the map resolution increases. More clumps and smaller clumps appear
above the density threshold at higher resolution and some massive
clumps disappear as they pick up fine-scale subregions below threshold.
On the other hand, the mass function slope is independent of resolution
in our model. Higher resolution extends the mass function to lower mass
and changes the total count of clouds for fixed mass above the density
threshold, but it leaves the mass function slope the same.

Physically, our model applies when only the dense parts of a cloud
complex are identified as clumps. These clumps can have highly
irregular shapes, like filaments, but they cannot include multiple
islands of threshold density gas separated by lower-density
inter-island gas. Our model would identify each island as a separate
clump. In Press \& Schechter (1974) and Hennebelle \& Chabrier (2008),
entire regions with islands and inter-island gas are counted as single
clumps if the average density of it all exceeds the threshold. In all
cases, the total mass in the mass function comes only from the parts of
the clumps that are above the density threshold; only the definition of
a clump and the counting of clumps differ. Which clump definition more
accurately applies to pre-stellar clumps depends on how observers
define their clumps and the masses of their clumps. While this may seem
like a small detail, the slope of the clump mass spectrum, $\alpha$,
depends on the slope of the power spectrum, $\beta$, in opposite ways
in these two cases.  If clump finding algorithms change their
identification of clumps as the spatial resolution increases, calling
what used to be a single clump at low resolution two separate clumps at
higher resolution, then our definition of clumps would apply.

\subsection{Projected Clouds}

Observations do not generally determine the 3D power spectrum of
density in a 3D cloud, but rather the 2D power spectrum of projected
density in a 3D cloud. This difference between the observed and the
intrinsic power spectra has to be investigated before we link the
observed power spectrum to a 3D clump mass function. Two-dimensional
power spectra were determined for the projected density distributions
and for thin slices in 3D fBm clouds. According to equation (28) in \citet*{laz}, the slope of the 2D power spectrum of
projected density, $\beta_{\rm thick}$, integrated along a thick path
through a 3D cloud, should be equal to the slope of the power spectrum
in the original 3D cloud, $\beta$. The slope of the power spectrum of a
thin slice in a 3D cloud, $\beta_{\rm thin}$, should differ by 1 from
the 3D slope: $\beta_{\rm thin}=\beta-1$. The power spectrum is
shallower for a thin slice than a projection because the thin slice has
more fine-scale structure. Figure \ref{psslope} shows these 2D power
spectra for a 3D fBm cloud with $\beta=2.8$. The thick path projection
through a 3D fBm cloud (top panel) was determined by integrating along
the $z$ axis from $z=1$ to $z=L$, which is the 3D box size. The thin
slice projection (bottom panel) was determined by using only the
$z=L/2$ position to make the 2D map. Figure \ref{psslope} indicates
that the 2D power spectra have approximately the slopes that are
expected. The projected density field has a 2D power spectrum slope
$\beta_{\rm thick}=-2.76$, and the 2D power spectrum of the thin slice
has a slope $\beta_{\rm thin}=1.92$. These are approximately equal to,
and 1 less than, respectively, the 3D slope of $\beta=2.8$.

This result for 2D indicates that observation of 2D power spectra for
projected or line-of-sight integrated clouds having slopes of about
$-2.8$ (Table 1) correspond to 3D power spectra inside those clouds
with about the same slope. Correspondingly, Figures \ref{beta2-82final}
and \ref{slopes} now suggest that for this slope, the mass function of
clumps has a slope equal to the Salpeter value, $\alpha=2.35$.  Thus
{\it the power-law portion of the stellar IMF can have its fundamental
origin in the geometric structure of clouds.} Section \ref{sect:disc}
discusses the implications of this point in more detail.

\begin{figure}
\epsfig{figure=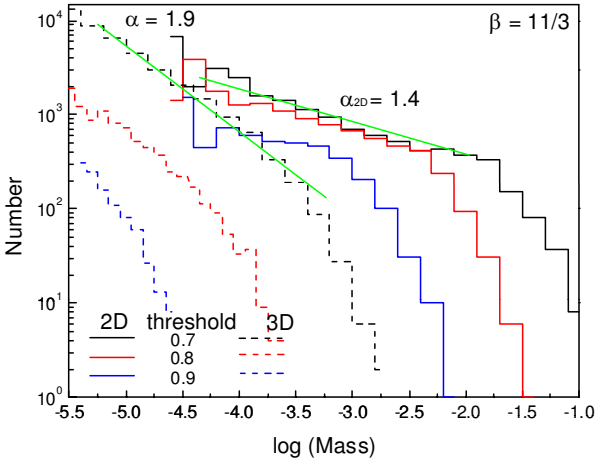,angle=0,scale=.45}\caption{Projected mass
functions of clumps in 3D fBm clouds obtained by integrating along the
$z$ axis. The mass spectra are shown by solid lines for thresholds of
$0.7$, $0.8$ and $0.9$. The mass functions of the corresponding 3D fBm
clouds are shown by dashed lines. The power spectrum slope is
$\beta=11/3$. Green lines are fitted to the 2D and 3D mass functions
for a density threshold of $0.7$.\label{projected}}
\end{figure}

Mass functions for clumps observed in projection are shown in Figure
\ref{projected}. These were made from the same 3D fBm clouds used for
Figure \ref{slopes}, but the search algorithm was done on 2D
projections of the clouds made by integrating along the $z$ axis. Mass
functions for the 3D fBm clumps shown previously are reproduced as
dashed lines. Three different density thresholds are considered with a
3D power spectrum slope $\beta=11/3$. This is the slope of the 3D power
spectrum for velocity in classical 3D incompressible Kolmogorov
turbulence.  The green lines in Figure \ref{projected} show fits for a
density threshold of $0.7$. As for the 3D mass functions, the slopes of
the low-mass parts of the 2D mass functions are about the same for all
density thresholds; the slopes of the upper mass parts are steeper, and
the upper mass limits are smaller, for higher density thresholds. The
slope of the 3D mass function is written in the Figure as $\alpha=1.9$.
This is also the value expected from Figure \ref{slopes} for
$\beta=11/3$. The slope of the projected 2D mass function is
$\alpha_{\rm 2D}=1.4$.  Evidently, the 2D mass function of the
projected 3D cloud is shallower than the 3D mass function in the
unprojected cloud by $\sim0.5$ in the slope. This implies that the 2D
mass function has proportionally more high mass clumps, presumably from
the blending of smaller clumps.

The 2D mass function of projected clumps is not necessarily the same as
the mass function of interstellar clouds observed with dust emission
integrated along a long line of sight. If there is blending on the line
of sight because each clump is relatively large in angle, as would
result from a low threshold density for the definition of a clump, then
this projection result might apply. However, if the density threshold
is very high and there is no blending for a cloud with a finite
line-of-sight path length, then the observed projected mass function
could be closer to the true 3D mass function.

\subsection{fBm clouds with log-normal density pdf's}

A second set of models was made with log-normal density pdfs, which are
exponentials of the previously discussed fBm clouds (Equation
\ref{eq:rho}; see also \citet*{E2002}). We considered three types of
density normalization. The first forced each log-normal cloud to have a
minimum density, $\rho_{\rm LN}$, of 0 and a maximum density of $e^1$.
This was done by selecting $\rho_0$ in equation (\ref{eq:rho}) to be
the maximum value of $\rho_{\rm fBm}$, designated as $\rho_{\rm
fBm,max}$. Each fBm cloud had a different $\rho_0=\rho_{\rm fBm,max}$.
Clumps were defined to be regions where the density $\rho_{\rm LN}$ was
larger than some fixed fraction, $f_{\rm clump}$, of the peak value of
$e^1$. The same fraction was used for all of the log-normal
distributions in the ensemble; different $f_{\rm clump}$ were used to
see how the mass distribution depended on the threshold density.

A second case of $\rho_{\rm LN}$ normalization had a minimum density of
0 and a different maximum density for each cloud. The normalization
$\rho_0$ was now taken to be the same for each cloud and equal to the
average of all the $\rho_{\rm fBm,max}$ from the 100 previous
ensembles, $\rho_0= <\rho_{\rm fBm,max}>$.  Thus the density maximum in
a log-normal cloud was $\rho_{\rm LN,max}= \exp (\rho_{\rm
fBm,max}/<\rho_{\rm fBm,max}>)$. The random numbers were the same as in
the first case. The density threshold for the definition of a clump was
set equal to a fraction $f_{\rm clump}$ of each density peak, i.e.,
$f_{\rm clump}\rho_{\rm LN,max}$. Because the density peaks are all
different in this normalization, the density thresholds are all
different too, although the density threshold fractions are the same.
The fractions are taken to be the same as in the first normalization
case, with a different $f_{\rm clump}$ for each ensemble of 100 clouds.

The third case of $\rho_{\rm LN}$ normalization had a minimum density
of 0 and a variable maximum $\rho_{\rm LN,max}$ defined in the same way
as in the second case, i.e., with the same $\rho_0= <\rho_{\rm
fBm,max}>$ for each cloud. Now the density threshold to define a clump
was taken to be the same for each cloud, not variable with the same
fraction of the peak.

The motivation for considering these different cases is our hypothesis
that cloud collapse to stars begins to get significant at a certain
characteristic density where the microscale processes change \citep{E2007}. We would like to know, for example, how the clump mass spectrum
depends on the peak gas density in a kpc-size region for a fixed
characteristic density (this is closest to case 3). We would also like
to know how it depends on the relative density of the collapse
threshold (close to case 2). If the absolute density for the onset of
collapse is the same everywhere, then the timescale for collapse would
be about the same and the star formation rate should scale as the first
power of the average ISM density. If, on the other hand, the relative
density for collapse is the same everywhere, then the timescale will
scale with the inverse square root of the average density and the star
formation rate will scale with the 1.5 power of the average density.

\begin{figure*}
\epsfig{figure=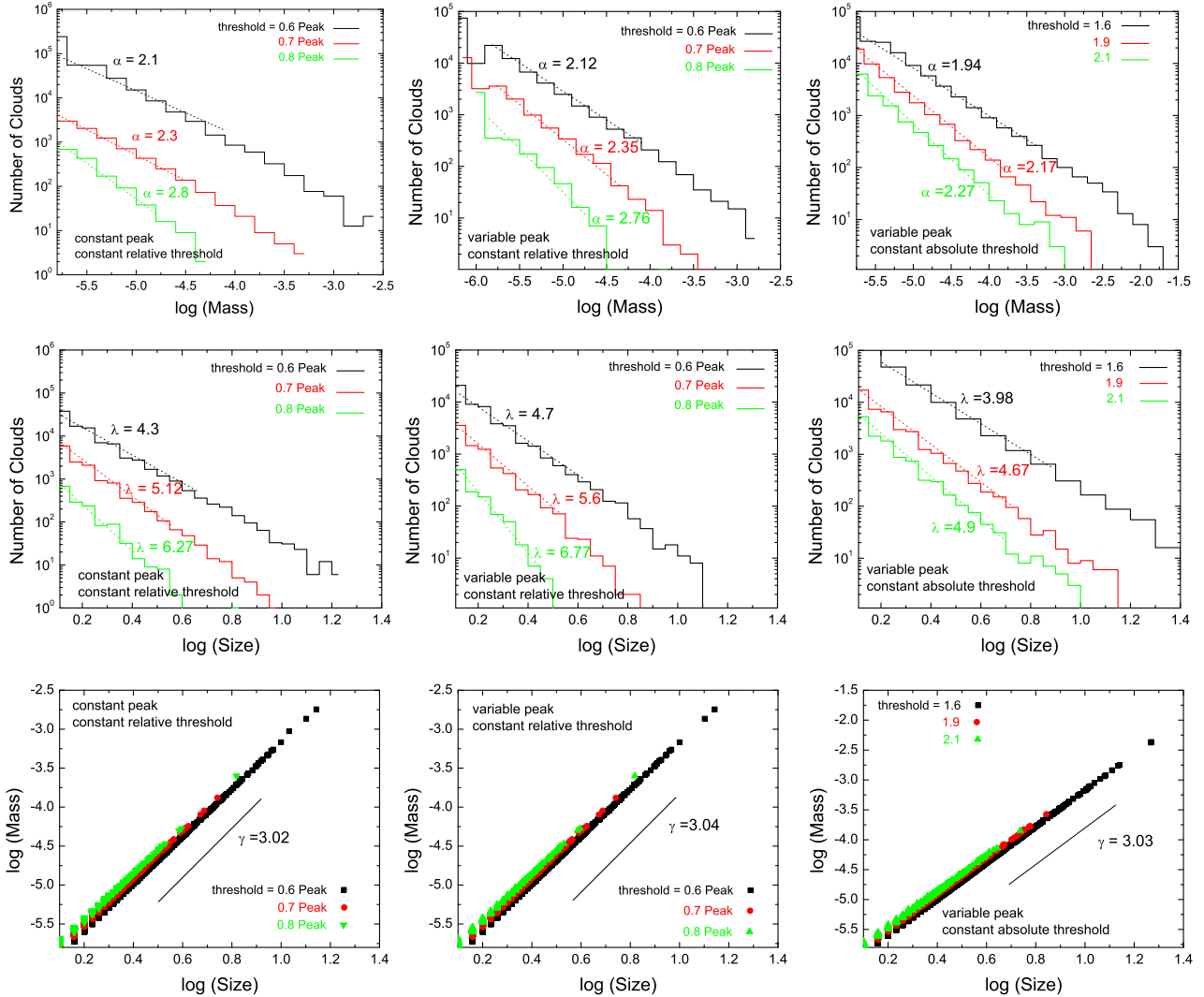,angle=0,scale=1.0} \caption{(Top) Clump mass
spectra of 3D fBm clouds with log-normal density pdfs and power law
slopes $\beta=2.8$. Three threshold values equal to $0.6$, $0.7$ and
$0.8$ are considered. (Middle) Size distributions of the clumps.
(Bottom) The mass-size relations. The three methods for cloud
normalization, corresponding to the three columns, are described
briefly in each panel and discussed in the text. \label{f1_mohsen}}
\end{figure*}

Figure \ref{f1_mohsen} shows the mass and size distributions and the
mass-size relation for the three different cases using $\beta=5/3$.
Each curve in the top two rows is a different clipping level, with
lower levels producing larger numbers of clumps. The clipping levels
were chosen to avoid excessive cloud merging at low levels and too few
clouds at high levels.  Each dot in the lower row represents a single
clump. Lower clipping levels produce lower mass clouds for the same
size.

The mass and size distributions are related to each other by the
mass-size correlation. For mass distribution $d n(M)/dM = M^{-\alpha}$,
size distribution $d\Lambda(R)/dR = R^{-\lambda}$, and mass-size
relation $M \propto R^{\gamma}$, we should have
$dn(M)/dM=(d\Lambda(R)/dR)(dR/dM)$, from which we get
$\alpha=1+(\lambda-1)/\gamma$. The bottom panels of Figure
\ref{f1_mohsen} indicate that $\gamma \sim 3.0$ for all cases, which
implies that the average density of clouds is nearly uniform when there
is a threshold density to define the clouds. This average density is
approximately 1.08 times the threshold density, depending a little on
$\beta$, because most the cloud volume is near the cloud edge where the
density equals this threshold.  Recall that clumps made from fBm clouds
are fractals with convoluted surfaces near the threshold density, so a
higher fraction of the mass is near this threshold density than in an
isothermal sphere.  For $\gamma=3.0$, $\alpha=1+0.33(\lambda-1)$. All
of our 3D results approximately follow this expression.

The mass functions and size functions in 3D resemble those in 1D and in
the regular fBm case in the sense that they have similar slopes at low
mass but then steepen to different degrees at high mass. The mass where
the steepening begins is lower for higher threshold densities.  This
trend is present for all three cases in Figure \ref{f1_mohsen}. The
average slope of the mass function is a blend of the common low-density
slope and the steepened high-density slope. This average steepens from
$-2.1$ to $-2.8$ as the relative threshold density increases from $0.6$
to $0.8$ in the first normalization case. The average slope steepens in
the second case from $-2.12$ to $-2.76$ and in the third case from
$-1.94$ to $-2.27$.

The steepening with increasing threshold density is lowest in the third
case, which is the one with a variable peak and a constant threshold
for clump definition. The variable relative density threshold that
results in this case effectively blends together the mass functions
from cases with higher and lower absolute thresholds.

\section{Discussion: How important is Cloud Structure to the IMF?}\label{sect:disc}

As a step toward understanding the IMF, we have investigated the mass
functions of clumps that exceed a threshold density in fractal Brownian
motion clouds.  We expected that mass functions will resemble volume
functions when there is a threshold density, and we verified that here.
We also found relations in 1D and 3D between the slopes of the power
laws for the mass functions and the power spectra, reminiscent of the
analytical results in Stutzki, et al. (1998) and Hennebelle \& Chabrier
(2008), although not exactly the same.  For the most realistic case, a
3D fBm cloud with a projected power spectrum similar to what is
observed in molecular clouds, the mass function of clumps denser than a
threshold is a power law with the Salpeter slope. This result does not
depend on the threshold density much, nor on the projection. Only the
maximum clump mass depends on the threshold density.  As a result of
this agreement between the clump mass spectrum and the stellar IMF for
realistic structure in interstellar clouds, we are led to believe that
the  power-law part of the IMF results in large part from cloud
structure  at a threshold density.  The low-mass part of the IMF
would then arise from an inefficiency of star formation below a certain
mass, like the thermal Jeans mass. We consider this in more detail
here.

The clump function is a reasonable starting point for studies of the
IMF, but the IMF cannot be explained so simply. If there is a fixed
density threshold \citep[e.g.,][]{johnstone,E2007}, then larger clumps with the same temperature
contain more Jeans masses and should fragment into more sub clumps and
stars. The assumed one-to-one correspondence between clump mass and
star mass is lost. This is a problem faced by observations of the pipe
nebula \citep{rathborne}: the clump mass distribution resembles
the IMF, but the largest clump contains many stars and breaks the
assumption of a one-to-one correspondence between clump mass and star
mass. On the other hand, if there is such a correspondence, i.e., if
the Jeans mass scales with the clump mass, then massive stars form in
lower density clumps or in the low-density periphery of clumps (in the
absence of temperature variations), and this positioning is contrary to
the observed mass segregation in young clusters. More physical
processes are required, such as heating.

An additional problem is that if larger clumps contain more Jeans
masses and form more stars, then there would be a stellar mass function
inside each clump, which could, in principle, differ from the clump
mass function. Only the sum of the IMFs inside all of the clumps has to
equal the total cloud IMF, which is the generally observed function.
There is virtually no systematic study of individual (local) IMFs
inside embedded-star clumps.

There are several ways in which the sum of individual clump IMFs can
give the observed IMF. One is for the IMF inside each clump (the
"internal clump IMF") to be the same as the cluster IMF, up to a
certain maximum star mass dependent on the clump mass, and for the
clump mass function to have a slope shallower than or equal to
$\alpha=2$ \citep*{EE2006}. This seems ruled out by the observation of
clump mass functions with steeper slopes, similar to that of the IMF.
A steeper clump function produces a summed IMF that is steeper than the
individual clump IMF (see below).

Another way to give the right summed IMF is for the internal clump IMF
to be a random sample of the cluster IMF up to arbitrarily high stellar
mass in each clump. Then the clump mass function does not enter into
the IMF. However, it is unrealistic to expect each clump to be able to
produce a star of arbitrarily large mass. With additional physical
processes, such as continued accretion from other clumps, random IMF
sampling might be possible, but not with static clumps as the only
reservoirs for internal clump IMFs.

In a third way, the internal clump IMF can be an arbitrary function of
the star-to-clump mass ratio. Then the summed IMF will have the same
power as the clump mass function in the power-law range, regardless of
the internal clump IMF. In this third case, the
universality of the IMF follows from the universality of cloud
structure.  The present study of fBm clump mass functions fits closest
to this third scenario. There need not be a one-to-one correspondence
between clump mass and star mass. Instead, there has to be an internal
IMF for each clump that is relative, i.e., the stellar masses scale up
or down with the clump mass.  This might be the case if competitive
accretion dominates the assembly of stars inside each clump, and if
heating, merging, ejection, and other processes contribute to the final
stellar IMF inside each clump, regardless of what that IMF is, as long
as all of the rates and masses are proportional to the clump mass.

These three scenarios are worth reviewing here.  We consider in all
cases a constant ratio of total star mass to clump mass in each clump,
so the total star mass in each clump is equal to some fraction
$\epsilon$ of the clump mass. We also consider for simplicity only the
power-law part of the IMF, ignoring the low-mass plateau. The clump
mass function is $dn_c(M_c)/dM_c=n_{c0}M_c^{-\alpha}$, as before.

For the first of the three scenarios, we consider a star mass function
inside each clump of $dn_s(M_s)/dM_s =n_{s0}M_s^{-1-x}$ for Salpeter
$x=1.35$. The integral of $M_s$ times the star mass function from a
fixed minimum stellar mass $M_{s,min}$ up to some very large stellar
mass has to equal the total stellar mass in the clump, which is
$\epsilon M_c$. That means $n_{s0}=A\epsilon M_{c}$ for
$A=(x-1)M_{s,min}^{x-1}$, and it means the IMF in a clump of mass $M_c$
is $n_s(M_s|M_c)=A\epsilon M_cM_s^{-1-x}$. The summed star mass
function for all of the clumps is the integral of $M_sn_s(M_s|M_{c})$
over the clump mass function from the minimum clump mass that is likely
to contain a star of mass $M_s$, which is
$M_{c,min}=xM_s^x/\left(A\epsilon\right)$, up to the maximum clump mass
in the cloud, $M_{c,max}$. The minimum clump mass comes from the
requirement that $\int_{M_s}^\infty n_{s0}M_s^{-1-x}dM_s=1$.  The
summed star mass is therefore
\begin{displaymath}
n_{sum}(M_s)=\int_{M_{c,min}}^{M_{c,max}} A\epsilon M_c M_s^{-1-x}
n_{c0}M_c^{-\alpha} dM_c
\end{displaymath}
\begin{equation}
 = {{A\epsilon n_{c0}M_s^{-1-x}}\over{2-\alpha}}
\left(M_{c,max}^{2-\alpha} -\left[ {{xM_s^x}\over
{A\epsilon}}\right]^{2-\alpha}\label{ns}
 \right)
\end{equation}
for $\alpha\ne2$. For $\alpha=2$, it is
\begin{equation}
n_{sum}(M_s)=A\epsilon n_{c0}M_s^{-1-x}\ln\left(A\epsilon
M_{c,max}/\left[ xM_s^x\right] \right).\end{equation} When
$\alpha\leq2$, the first term in the parenthesis of equation (\ref{ns})
dominates the second and the summed stellar mass function is $\propto
M_s^{-1-x}$, which is the same as the individual clump IMF. If
$\alpha=2$, the same is true approximately, with only a logarithmic
deviation. If $\alpha>2$, then the second term in the parentheses of
equation (\ref{ns}) dominates the first, and the summed IMF is $\propto
M^{-1-x(\alpha-1)}$. This summed IMF can be much steeper than the
Salpeter function if $\alpha>2$ \citep{kroupa}. Because dense
clumps are observed to have $\alpha>2$, individual clump IMFs have to
be shallower than the whole-cluster IMF, which is the Salpeter
function.

For the second of the three scenarios, a star of any mass is supposed
to be able to form in a clump of any mass, in which case $M_{c,min}$ is
constant in the integral of equation \ref{ns}, independent of $M$. Then
the summed IMF is trivially proportional to the individual clump IMF,
$M_s^{-1-x}$.

The third scenario allows a variety of stellar mass functions inside
each clump and always gives a summed stellar function with a slope that
is equal to the slope of the clump mass function.  To demonstrate this,
we assume the internal clump IMF is a distribution function, $P(f)df$,
of the ratio of star mass to clump mass, $f=M_s/M_c$ for $f$ between
$f_{min}$ and $f_{max}$. For simplicity, $P(f)=P_0f^{-p}$. Then the
summed stellar mass function is
\begin{displaymath}
n_{sum}(M_s)=\int_{f_{min}}^{f_{max}} P(f) n_c(M_c)df
\end{displaymath}
\begin{equation}
=\int_{M_{c,min}}^{M_{c,max}} P_0 M_s^{1-p} M_c^{p-\alpha-2} dM_c ,
\end{equation}
which integrates to
\begin{equation}
n_{sum}(M_s)={{P_0M_s^{1-p}}\over{(p-\alpha-1)}}
\left(M_{c,max}^{p-\alpha-1}-M_{c,min}^{p-\alpha-1} \right).
\end{equation}
If clumps larger than a certain mass $M_J$ fragment into stars, then
$M_{c,min}=MAX(M_J,M_s/f_{max})$. Also, $M_{c,max}=M_sf_{min}$.

Now consider clump masses larger than the fragmentation limit, i.e.
$M_s/f_{max}>M_J$. Then $M_{c,min}=M_s/f_{max}$, and the stellar mass
function is
\begin{equation}
n_{sum}(M_s)={{P_0M_s^{-\alpha}}\over{(p-\alpha-1)}}
\left(f_{min}^{(1+\alpha+p)}-f_{max}^{(1+\alpha-p)}\right).
\end{equation}
This IMF is proportional to $M_s^{-\alpha}$, which is the same as the
clump mass function with a slope that is independent of the individual
clump IMF slope $p$. Below the fragmentation limit but not as low in
stellar mass as the minimum fragment size, we have $M_{c,min}=M_J$ and
also $M_s>M_Jf_{min}$. Then
\begin{displaymath}
n_{sum}(M_s)={{P_0M_s^{1-p}M_J^{(p-\alpha-1)}}\over{(1+\alpha-p)}}
\end{displaymath}
\begin{equation}
\times \left(1-\left[{{M_Jf_{min}} \over{M_s}}\right]^{(1+\alpha-p)}\right),
\end{equation}
where the second term in the brackets is less than the first for
$1+\alpha>p$. In this case, the summed IMF is proportional to
$M_s^{1-p}$, which is the function $P(p)$ in intervals of $\log P$.

This result for the third scenario implies that if clumps need to
contain at least one Jeans mass before they fragment into a little
sub-cluster, and if the Jeans mass, $M_J$, is about the same for all
clumps (i.e., for a constant temperature and threshold density), then
the summed IMF above $M_Jf_{max}$, which is the largest star that comes
from an $M_J$ piece, is a power law with the same slope as the clump
mass function. Below $M_Jf_{max}$, the summed IMF is the internal clump
mass function. If $p\sim2$, then the IMF plateau may be explained as
the part below the smallest unstable clump. It would be composed of
stars coming from that smallest unstable clump, and from other clumps
with slightly larger masses \citep{Elm2000}.

What type of model will produce a relative stellar mass function,
$p(f)$, inside each clump, rather than an absolute stellar mass
function inside each clump? Such a function would seem to result from
self-similar processes that scale to the clump mass, and that would
include protostellar accretion of a gas mass proportional to the clump
mass, and protostellar interactions, including possibly coalescence. As
all of this occurs in clumps more massive than $M_J$, the motions would
be transonic and $M_J$ itself would be unconnected with the final star
mass. Note that in this model, the number of fragments in a clump would
not be proportional to the number of $M_J$ in the clump, because that
would give stellar masses that are centered around $M_J$ rather than
stellar masses that scale with the clump mass. In this interpretation,
$M_J$ enters the final IMF in its establishment of a minimum
fragmentable clump mass at the IMF plateau.  Also in this model,
massive clumps would not produce more stars than low-mass clumps with
the same average stellar mass, but they would produce the same number
of stars as the low-mass clumps and a higher mass for each star. Thus
the third model is like assuming each clump produces stars with masses
proportional to the clump mass. It is effectively the same as assuming
a one-to-one correspondence between clump mass and star mass, even
though each clump forms several stars with a local, and even arbitrary,
relative IMF.

\section{Conclusions}

Interstellar clouds have a power spectrum of column density that is a
power law with a negative slope in the range from $\sim2.7$ to 3. The
underlying structure is not a clump-interclump medium but a continuum
of densities with structures in the form of sheets, filaments, and
clumps. Surveys that are sensitive to column density or density view
these clouds as collections of discrete clumps, which are the densest
parts of the continuum.  In many models, star formation also selects
the densest parts of a cloud. Here we investigated a model for the IMF
based on this principle. We determined clump mass spectra above
threshold densities for clouds with power-law power spectra. The
results showed a dependence of the clump mass spectrum on the slope of
the power spectrum, as expected from analytic theory, and a trend
toward decreasing maximum mass with increasing relative threshold
density. For realistic power spectra, an IMF-like slope results for the
derived clump spectra. At a very basic level, this result supports
numerous models of star formation where the intermediate-to-high mass
part of the IMF has its basic form from cloud geometry.

The IMF has to be more complicated than this, however.  If there is a
characteristic density for star formation and a wide range of clump
masses above that density, as is the case for clouds with power-law
power spectra, then larger clumps contain more unstable sub-clumps and
should form more stars or more massive stars. If they just form more
stars with the same characteristic mass, e.g., the thermal Jeans mass,
then the assumed connection between clump mass and star mass is lost,
as is the connection between the clump mass function and the IMF. The
IMF needs a different origin in that case. If, on the other hand, the
average star mass inside a clump scales with the clump mass, then the
clump mass spectrum can be the same as the IMF in the power-law part.
Such scaling allows for the interesting possibility that clumps have
internal IMFs that are different from observed IMFs in whole clusters,
which is the sum from many clumps. Internal IMFs inside clumps have not
been investigated much so this possibility remains open.

Competitive accretion and clump coalescence play an important role in
forming the IMF, as shown by many numerical simulations. These
processes usually operate along with cloud fragmentation, so it is
unclear whether the bulk of the IMF power law still comes from
geometry. That is, the basic $dn/dM\propto M^{-2}$ power law that
arises in so many fragmentation models is already very close to the IMF
(the Salpeter slope is $=-2.35$ in this convention). Accretion, stellar
winds, coalescence, and other microphysics need only modify this
initial function by a small amount, giving a slight bias toward lower
mass objects. Experiments of competitive accretion in clouds with
different density power spectra or in uniform clouds might reveal the
overall importance of cloud geometry in the IMF.

\section*{Acknowledgments}
We are grateful to the  referee, Patrick Hennebelle, whose detailed and
careful comments helped to improve the quality of this paper. MS is
happy to acknowledge the hospitality of the staff and Axel Brandenburg
at NORDITA where parts of this work were done during a research visitor
program.

\bibliographystyle{mn2e}
\bibliography{referenceIMF}

\end{document}